%Paper: hep-ph/9410282
%From: tomislav@hepth.cornell.edu
%Date: Thu, 13 Oct 94 23:52 BST
%Date (revised): Wed, 25 Oct 95 11:54:23 +0100
%Date (revised): Mon, 30 Oct 1995 14:19:30 -0500 (EST)
%Date (revised): Tue, 31 Oct 1995 14:44:29 -0500 (EST)

%%%%%%%%%%%%%%%%%%  tex macros for harvard preprints, cm version %%%%%%%

%                      P. Ginsparg, P. Nelson
%                      last updated 4/88
%                      nonharvard users say "b" in response
%                      to query, and only plain.tex is needed
%
\newbox\leftpage \newdimen\fullhsize \newdimen\hstitle \newdimen\hsbody
\tolerance=1000\hfuzz=2pt
\def\bigans{b }
%\message{ big or little (b/l)? }\read-1 to\answ
\def\answ{b } % defined to make possible automated postscript creation
\ifx\answ\bigans\message{(This will come out unreduced.}
%\magnification=1200\baselineskip=16pt plus 2pt minus 1pt
\magnification=1200\baselineskip=18pt plus 2pt minus 1pt
%\magnification=1200\baselineskip=32pt plus 2pt minus 1pt
\hsbody=\hsize \hstitle=\hsize %take default values for unreduced format
\else\def\apans{l }\message{ lyman or hepl (l/h) (lowercase!) ? }
\read-1 to \apansw\message{(This will be reduced.}
\let\lr=L
\magnification=1000\baselineskip=16pt plus 2pt minus 1pt
\voffset=-.31truein\vsize=7truein
\hstitle=8truein\hsbody=4.75truein\fullhsize=10truein\hsize=\hsbody
%	Send local landscape command to laserprinter
\ifx\apansw\apans\special{ps: landscape}\hoffset=-.59truein% apple lw
  \else\hoffset=.05truein\fi% qms laserprinter
\output={\ifnum\pageno=0 %%% This is the HUTP version
  \shipout\vbox{\hbox to \fullhsize{\hfill\pagebody\hfill}}\advancepageno
  \else
  \almostshipout{\leftline{\vbox{\pagebody\makefootline}}}\advancepageno
  \fi}
\def\almostshipout#1{\if L\lr \count1=1
      \global\setbox\leftpage=#1 \global\let\lr=R
  \else \count1=2
    \shipout\vbox{\ifx\apansw\apans\special{ps: landscape}\fi %satisfies dvips
      \hbox to\fullhsize{\box\leftpage\hfil#1}}  \global\let\lr=L\fi}
\fi
%
%---------------------------------------------------------------------
\catcode`\@=11 % This allows us to modify PLAIN macros.
\newcount\yearltd\yearltd=\year\advance\yearltd by -1900

\def\Title#1#2{\nopagenumbers\abstractfont\hsize=\hstitle\rightline{#1}%
\vskip 1in\centerline{\titlefont #2}\abstractfont\vskip .5in\pageno=0}
\def\Date#1{\vfill\leftline{#1}\tenpoint\supereject\global\hsize=\hsbody%
\footline={\hss\tenrm\folio\hss}}% restores pagenumbers
\def\draftmode{\def\draftdate{{\rm preliminary draft:
\number\month/\number\day/\number\yearltd\ \ \hourmin}}%
\headline={\hfil\draftdate}\writelabels\baselineskip=20pt plus 2pt minus 2pt
{\count255=\time\divide\count255 by 60 \xdef\hourmin{\number\count255}
	\multiply\count255 by-60\advance\count255 by\time
   \xdef\hourmin{\hourmin:\ifnum\count255<10 0\fi\the\count255}}}
%use this one instead of \Date on the preliminary draft
%it puts the current date on each page in big mode

%use \nolabels to get rid of eqn and ref labels in draft mode
\def\nolabels{\def\eqnlabel##1{}\def\eqlabel##1{}\def\reflabel##1{}}
\def\writelabels{\def\eqnlabel##1{%
{\escapechar=` \hfill\rlap{\hskip.09in\string##1}}}%
\def\eqlabel##1{{\escapechar=` \rlap{\hskip.09in\string##1}}}%
\def\reflabel##1{\noexpand\llap{\string\string\string##1\hskip.31in}}}
\nolabels
%
% tagged sec numbers
\global\newcount\secno \global\secno=0
\global\newcount\meqno \global\meqno=1
\def\newsec#1{\global\advance\secno by1
\xdef\secsym{\the\secno.}\global\meqno=1
%\ifx\answ\bigans \vfill\eject\else
\bigbreak\bigskip%\fi% (combination \goodbreak\bigskip\bigskip)
\noindent{\bf\the\secno. #1}\par\nobreak\medskip\nobreak}
\xdef\secsym{}
%
\def\appendix#1#2{\global\meqno=1\xdef\secsym{\hbox{#1.}}\bigbreak\bigskip
\noindent{\bf Appendix #1. #2}\par\nobreak\medskip\nobreak}
%
%       \eqn\label{a+b=c}	gives a displayed equation with number
%				chosen consecutively within sections.
%     \eqnn and \eqna define labels in advance
%
\def\eqnn#1{\xdef #1{(\secsym\the\meqno)}%
\global\advance\meqno by1\eqnlabel#1}
\def\eqna#1{\xdef #1##1{\hbox{$(\secsym\the\meqno##1)$}}%
\global\advance\meqno by1\eqnlabel{#1$\{\}$}}
\def\eqn#1#2{\xdef #1{(\secsym\the\meqno)}\global\advance\meqno by1%
$$#2\eqno#1\eqlabel#1$$}
%
%			 footnotes
\newskip\footskip\footskip14pt plus 1pt minus 1pt %sets footnote baselineskip
\def\f@@t{\baselineskip\footskip\bgroup\aftergroup\@foot\let\next}
\setbox\strutbox=\hbox{\vrule height9.5pt depth4.5pt width0pt}
\global\newcount\ftno \global\ftno=0
\def\foot{\global\advance\ftno by1\footnote{$^{\the\ftno}$}}
%
%     \ref\label{text}
% generates a number, assigns it to \label, generates an entry.
% To list the refs on a separate page,  \listrefs
%
\global\newcount\refno \global\refno=1
\newwrite\rfile
\def\ref{[\the\refno]\nref}
\def\nref#1{\xdef#1{[\the\refno]}\ifnum\refno=1\immediate
\openout\rfile=refs.tmp\fi\global\advance\refno by1\chardef\wfile=\rfile
\immediate\write\rfile{\noexpand\item{#1\ }\reflabel{#1}\pctsign}\findarg}
%	horrible hack to sidestep tex \write limitation
\def\findarg#1#{\begingroup\obeylines\newlinechar=`\^^M\pass@rg}
{\obeylines\gdef\pass@rg#1{\writ@line\relax #1^^M\hbox{}^^M}%
\gdef\writ@line#1^^M{\expandafter\toks0\expandafter{\striprel@x #1}%
\edef\next{\the\toks0}\ifx\next\em@rk\let\next=\endgroup\else\ifx\next\empty%
\else\immediate\write\wfile{\the\toks0}\fi\let\next=\writ@line\fi\next\relax}}
\def\striprel@x#1{} \def\em@rk{\hbox{}} {\catcode`\%=12\xdef\pctsign{%}}
\def\semi{;\hfil\break}
\def\addref#1{\immediate\write\rfile{\noexpand\item{}#1}} %now unnecessary
\def\listrefs{
\vfill\eject\immediate\closeout\rfile%\parindent=20pt
\baselineskip=24pt\centerline{{\bf References}}\bigskip{\frenchspacing%
\escapechar=` \input refs.tmp\vfill\eject}\nonfrenchspacing}
\def\startrefs#1{\immediate\openout\rfile=refs.tmp\refno=#1}
\def\figures{\centerline{{\bf Figure Captions}}\medskip\parindent=40pt}
\def\fig#1#2{\medskip\item{Figure ~#1:  }#2}
\catcode`\@=12 % at signs are no longer letters
%
%	Unpleasantness in calling in abstract and title fonts
\ifx\answ\bigans
\font\titlerm=cmr10 scaled\magstep3 \font\titlerms=cmr7 scaled\magstep3
\font\titlermss=cmr5 scaled\magstep3 \font\titlei=cmmi10 scaled\magstep3
\font\titleis=cmmi7 scaled\magstep3 \font\titleiss=cmmi5 scaled\magstep3
\font\titlesy=cmsy10 scaled\magstep3 \font\titlesys=cmsy7 scaled\magstep3
\font\titlesyss=cmsy5 scaled\magstep3 \font\titleit=cmti10 scaled\magstep3
\else
\font\titlerm=cmr10 scaled\magstep4 \font\titlerms=cmr7 scaled\magstep4
\font\titlermss=cmr5 scaled\magstep4 \font\titlei=cmmi10 scaled\magstep4
\font\titleis=cmmi7 scaled\magstep4 \font\titleiss=cmmi5 scaled\magstep4
\font\titlesy=cmsy10 scaled\magstep4 \font\titlesys=cmsy7 scaled\magstep4
\font\titlesyss=cmsy5 scaled\magstep4 \font\titleit=cmti10 scaled\magstep4
\font\absrm=cmr10 scaled\magstep1 \font\absrms=cmr7 scaled\magstep1
\font\absrmss=cmr5 scaled\magstep1 \font\absi=cmmi10 scaled\magstep1
\font\absis=cmmi7 scaled\magstep1 \font\absiss=cmmi5 scaled\magstep1
\font\abssy=cmsy10 scaled\magstep1 \font\abssys=cmsy7 scaled\magstep1
\font\abssyss=cmsy5 scaled\magstep1 \font\absbf=cmbx10 scaled\magstep1
\skewchar\absi='177 \skewchar\absis='177 \skewchar\absiss='177
\skewchar\abssy='60 \skewchar\abssys='60 \skewchar\abssyss='60
\fi
\skewchar\titlei='177 \skewchar\titleis='177 \skewchar\titleiss='177
\skewchar\titlesy='60 \skewchar\titlesys='60 \skewchar\titlesyss='60
\def\titlefont{\def\rm{\fam0\titlerm}% switch to title font
\textfont0=\titlerm \scriptfont0=\titlerms \scriptscriptfont0=\titlermss
\textfont1=\titlei \scriptfont1=\titleis \scriptscriptfont1=\titleiss
\textfont2=\titlesy \scriptfont2=\titlesys \scriptscriptfont2=\titlesyss
\textfont\itfam=\titleit \def\it{\fam\itfam\titleit} \rm}
\ifx\answ\bigans\def\abstractfont{\tenpoint}\else
\def\abstractfont{\def\rm{\fam0\absrm}% switch to abstract font
\textfont0=\absrm \scriptfont0=\absrms \scriptscriptfont0=\absrmss
\textfont1=\absi \scriptfont1=\absis \scriptscriptfont1=\absiss
\textfont2=\abssy \scriptfont2=\abssys \scriptscriptfont2=\abssyss
\textfont\itfam=\bigit \def\it{\fam\itfam\bigit}
\textfont\bffam=\absbf \def\bf{\fam\bffam\absbf} \rm} \fi
\def\tenpoint{\def\rm{\fam0\tenrm}% switch back to 10-point type
\textfont0=\tenrm \scriptfont0=\sevenrm \scriptscriptfont0=\fiverm
\textfont1=\teni  \scriptfont1=\seveni  \scriptscriptfont1=\fivei
\textfont2=\tensy \scriptfont2=\sevensy \scriptscriptfont2=\fivesy
\textfont\itfam=\tenit \def\it{\fam\itfam\tenit}
\textfont\bffam=\tenbf \def\bf{\fam\bffam\tenbf} \rm}
%
%---------------------------------------------------------------------
%
\def\noblackbox{\overfullrule=0pt}
\hyphenation{anom-aly anom-alies coun-ter-term coun-ter-terms}
\def\inv{^{\raise.15ex\hbox{${\scriptscriptstyle -}$}\kern-.05em 1}}
\def\dup{^{\vphantom{1}}}
\def\Dsl{\,\raise.15ex\hbox{/}\mkern-13.5mu D} %this one can be subscripted
\def\dsl{\raise.15ex\hbox{/}\kern-.57em\partial}
\def\del{\partial}
\def\Psl{\dsl}
\def\tr{{\rm tr}} \def\Tr{{\rm Tr}}
\font\bigit=cmti10 scaled \magstep1
\def\biglie{\hbox{\bigit\$}} %pound sterling
\def\lspace{\ifx\answ\bigans{}\else\qquad\fi}
\def\lbspace{\ifx\answ\bigans{}\else\hskip-.2in\fi} % $$\lbspace...$$
\def\boxeqn#1{\vcenter{\vbox{\hrule\hbox{\vrule\kern3pt\vbox{\kern3pt
	\hbox{${\displaystyle #1}$}\kern3pt}\kern3pt\vrule}\hrule}}}
\def\mbox#1#2{\vcenter{\hrule \hbox{\vrule height#2in
		\kern#1in \vrule} \hrule}}  %e.g. \mbox{.1}{.1}
%	matters of taste
%\def\tilde{\widetilde} \def\bar{\overline} \def\hat{\widehat}
%
% some sample definitions
\def\CAG{{\cal A/\cal G}}   %curly letters
\def\CA{{\cal A}} \def\CC{{\cal C}} \def\CF{{\cal F}} \def\CG{{\cal G}}
\def\CL{{\cal L}} \def\CH{{\cal H}} \def\CI{{\cal I}} \def\CU{{\cal U}}
\def\CB{{\cal B}} \def\CR{{\cal R}} \def\CD{{\cal D}} \def\CT{{\cal T}}
\def\e#1{{\rm e}^{^{\textstyle#1}}}
\def\grad#1{\,\nabla\!_{{#1}}\,}
\def\gradgrad#1#2{\,\nabla\!_{{#1}}\nabla\!_{{#2}}\,}
\def\ph{\varphi}
\def\psibar{\overline\psi}
\def\om#1#2{\omega^{#1}{}_{#2}}
\def\vev#1{\langle #1 \rangle}
\def\lform{\hbox{$\sqcup$}\llap{\hbox{$\sqcap$}}}
\def\darr#1{\raise1.5ex\hbox{$\leftrightarrow$}\mkern-16.5mu #1}
\def\lie{\hbox{\it\$}} %pound sterling
\def\ha{{1\over2}}
\def\half{{\textstyle{1\over2}}} %puts a small half in a displayed eqn
\def\roughly#1{\raise.3ex\hbox{$#1$\kern-.75em\lower1ex\hbox{$\sim$}}}

\Title{PUPT-94-1496, IN 94021, hep-ph/9410282}
{\vbox{\centerline{Non-local Electroweak Baryogenesis}
	\vskip2pt\centerline{Part II : The Classical Regime}}}

%   \footnote{}{*optional footnote on title}
\baselineskip 18pt
\centerline{{\bf Michael Joyce}\footnote{$^\dagger$}{Current e-mail:
joyce@nxth01.cern.ch, tomislav@hepth.cornell.edu,\hfil\break
 neil@puhep1.princeton.edu}}
\centerline{{\bf Tomislav Prokopec}{$^\dagger$}}
\centerline{and}
\centerline{{\bf Neil Turok}{$^\dagger$}}
\centerline{Joseph Henry Laboratories }
\centerline{Princeton University}
\centerline{Princeton, NJ 08544.}
\vskip .2in
\centerline{\bf Abstract}
\baselineskip=12pt
\noindent
We investigate baryogenesis at a first order electroweak phase
transition in the presence of  a CP violating condensate on the
bubble walls, in the regime in which the bubble walls are `thick',
in the sense that fermions interact with the plasma many times as
the bubble wall passes. Such a condensate is present in multi-Higgs
extensions of the standard model and may be formed via an instability
in the minimal standard model. We concentrate on particles with
typical thermal energies in the plasma, whose interactions with
the wall are accurately described by the WKB approximation,
in which a classical chiral force is evident. The deviations from
thermal equilibrium produced by motion of the wall are then treated
using a classical Boltzmann equation which we solve in a fluid
approximation. From the resulting equations we find two effects
important for baryogenesis: (i) a classical chiral force term due
to the $CP$ violating background, and (ii) a term arising from
hypercharge violating interactions which are pushed out of
equilibrium by the background{field. Provided the wall propagates
slower than the speed of sound, both terms lead to the diffusion of
a chiral asymmetry in front of the wall. This can produce a baryon
asymmetry of the observed magnitude for typical wall velocities
and thicknesses.

\Date{Revised: October 1995} %replace this line by \draft  for preliminary
%versions
	     %or specify \draftmode at some point
%\draft
\eject
\baselineskip 12pt

\centerline { \bf 1. Introduction}

In this  paper we  present a detailed
discussion of electroweak baryogenesis induced by a CP violating
condensate field on  `thick'
bubble walls during a first order electroweak
phase transition. `Thick' in this context, and as we will
see more precisely, means that
the mean free time for a fermion propagating in the plasma
is short compared to the time
taken for the wall to pass. In this case
one  expects  that the non-local quantum reflection
effects, which such a CP violating condensate has been
previously shown to produce
\ref\cknct{A. Cohen, D.  Kaplan and A.
 Nelson, Nuc. Phys. {\bf B373}, 453  (1992);
A. Cohen, D. Kaplan and A.  Nelson, Phys. Lett. {\bf B294}
(1992) 57. },\ref\JPTlept{
M. Joyce, T. Prokopec and N. Turok, Phys. Lett. {\bf B338}
(1994) 269.},\ref\funakubo{K. Funakubo, A. Kakuto, S. Otsuki, K. Takenaga
and F. Toyoda,
Phys. Rev. {\bf D50}, 1105 (1994);
preprints hep-ph/9405422, hep-ph/9407207, hep-ph/9503495. },
will be suppressed due to scattering.
Instead we look for  local  classical effects which
can produce a driving force for baryogenesis.
A classical treatment
has, as we shall see, many advantages in that there is a
systematic  framework (a Boltzmann transport equation)
within which to compute the nonequilibrium effects
in which we are interested.
A shorter version of this work has already appeared \ref\JPTfprl{
M. Joyce, T. Prokopec and N. Turok, Phys. Rev. Lett.{\bf 75},1695 (1994).}.

There is still considerable uncertainty as to what the
relevant bubble wall thickness and speed are.
Calculations are difficult
\ref\wall{N. Turok, Phys. Rev. Lett. {\bf 68}, 1803 (1992);
M. Dine, R. Leigh, P. Huet, A. Linde and D. Linde,
Phys. Rev. {\bf D46}, 550 (1992);
B-H. Liu, L. McLerran and N. Turok, Phys. Rev. {\bf D46}, 2668 (1992);
T. Prokopec and G. Moore, Phys. Rev. Lett. {\bf 75},777 (1995),
; PUP-TH-1544, LANCS-TH/9517 (1995)} and  strongly dependent
upon the still poorly determined effective potential.
A recent detailed study by one of us (T.P.) and G. Moore \wall\
using some of the methods developed in this paper
indicates, within the minimal standard model, for Higgs masses of order
30-70 GeV, and ignoring
possible nonperturbative effects, a wall velocity
$\sim 0.4$ and a  wall thickness of order $ \sim 25 T^{-1}$.
Thus typically
quarks interact very many times  via gluon exchange processes as they
cross the wall. If this is indeed the relevant regime,
then for top quarks at least
(the most obvious mediator of electroweak baryogenesis since they
couple most strongly
to the bubble wall if one has  standard model-like Yukawa couplings),
the particle-wall problem cannot be treated
without including the effects of strong (QCD) scattering from the plasma.

In an accompanying paper
\ref\JPTlonga{M. Joyce, T. Prokopec and N. Turok, Princeton
preprint PUPT-1495 (1994).}
we have introduced
the essential ideas motivating the calculations which we undertake
here.
We showed there that the Lagrangian for a fermion propagating
in the background of a bubble wall in a two Higgs doublet
extension
of the standard model can be written
\eqn\lagrangian{
{\cal L}=\overline{\Psi}\gamma^\mu i (\partial_\mu
- i g_A
\tilde{Z}_\mu\gamma^5)\Psi
- m \overline{\Psi}\Psi
}
where $g_A \tilde{Z}_\mu = g_A Z_\mu^{GI}
-{1\over 2} {v_2^2 \over v_1^2 +v_2^2} \partial_\mu \theta$,
$g_A= \pm {1\over 4} g$, $g=\sqrt{g_1^2+g_2^2}$ \JPTlonga.
 The $+$ sign is for up-type quarks  and (left-handed) neutrinos,
 the $-$ sign for down-type quarks and charged leptons.
$g_1$ and $g_2$ are the gauge couplings
of the $SU(2)$ and $U(1)$ gauge fields;
$v_1$ and $v_2$ the magnitudes of the vevs
of the two Higgs doublets, the first of which is
taken to couple to the fermions through Yukawa terms.
The two contributions to $\tilde{Z}_\mu$ come from
the $CP$-odd scalar
field $\theta$ which is the relative phase of the two
Higgs fields,
$\varphi^\dagger_2 \varphi_1 = R e^{i\theta}$, and
the $Z_{\mu}$ condensate discussed in
\ref\NasserTurok{S. Nasser
and N. Turok, Princeton preprint, PUPT-1456 (1994).}
which may be present even in the minimal theory.
All the vector couplings to $\tilde{Z}$ are
removed by using the remaining unbroken vector
symmetries to remove the pure gauge $\tilde{Z}$
(the $Z$ condensate piece is also pure gauge if we
treat the wall as planar, and assume it has reached a stationary
state in which the Higgs and gauge fields are functions of $z-v_wt$).
When the vevs vanish the remaining
axial $\tilde{Z}$ is a pure gauge field
(for both fermions and Higgs fields,  with charges
$g_A=0$ and $g_A=-{1 \over 2}g$ for the charged and neutral
Higgs components respectively, since
$g_A=\left [{1 \over 2}(T_3 - Y)+{1 \over 4}(B - L)\right ]g$ \JPTlonga\ ).

However, on the bubble wall
this pure gauge field has as we shall see
very tangible $CP$ violating effects,
even on particles with typical thermal energies.
The axial gauge field condensate $\tilde{Z}$ formed on
the bubble walls at the first order phase transition
violates CP spontaneously.
$\tilde{Z}^o$ is odd under CP, and
the spatial components $\tilde{Z}^i$ are  CP even so that
a bubble on which the spatial vector $\tilde{Z}^i$  points out
everywhere is mapped under CP to one on which it points in.
(Thus in the latter case it is actually the gradient $Z'$
which violates CP).

The departure from thermal equilibrium
which this brings about as the bubble wall moves through the
plasma will source baryogenesis.  In the companion paper
\JPTlonga\ we discussed
the case originally investigated by Cohen, Kaplan
and Nelson (CKN) \cknct,
when the fermion is treated as free on the bubble wall. The
conditions for the validity of this treatment are discussed
in \JPTlonga, and roughly require that the wall thickness $L$
is much less than the mean free path of the fermion $\lambda_f$.
In this paper we consider the perturbations in the plasma
produced by the CP violating background in the possibly
more realistic regime
of wall thickness for which the fermions interact
frequently on the bubble wall, a condition which will be
specified more precisely in the course of our treatment.

As we have discussed in \JPTlonga\ one  might expect that the
inclusion of interactions would wipe out any interesting CP violation
if  the effect
is a non-local quantum mechanical one. This is
precisely how the thin wall limit has been understood
\cknct, \JPTlept, \funakubo.
However as we
noted in \JPTlonga\ the WKB limit is not as trivial as it
appeared when viewed simply
in terms of reflection coefficients for monotonic wall ansatzes.
The fact that the dynamics of  WKB particles are non-trivial
- in particular that WKB particles propagate like particles in
a classical CP violating potential - suggests that there may be
interesting effects which do survive when the scattering
on the wall is included.

Secondly, as noted originally by CKN
\ref\cknsba{A. Cohen, D. Kaplan, and A. Nelson,
Phys. Lett. {\bf B263}, 86 (1991).},
such a CP violating background perturbs the energy levels of particles
and
can push processes out of equilibrium locally. The original form
of this `spontaneous' baryogenesis took this perturbation to the
energy to be modeled by a fermionic hypercharge potential,
and calculated the resultant chemical potential driving baryon
number violation by imposing constraints on exactly conserved quantum
numbers. Both of these aspects of the calculation have been criticized.
Dine and Thomas
\ref\dt{
M. Dine and S. Thomas, Phys. Lett. {\bf B328} (1994) 73.}
pointed out that the fermionic hypercharge potential cannot be
appropriate as the effect does not vanish as the Higgs
vev vanishes.
Subsequently, we pointed out
\ref\JPTtran{M. Joyce, T. Prokopec and N.
Turok, ``Constraints and Transport in Electroweak Baryogenesis'',
Phys. Lett. {\bf B393} (1994) 312.}
that the imposition of the constraints neglects
transport processes which tend to restore the region to a
local thermal equilibrium in which there is no baryon number violation.
The treatment we will present here will take account of both
these criticisms and show that the essential effect does survive
and can also, when transport is taken into account, source
significant perturbations in front of the bubble wall
\ref\JSintra{M. Joyce, hep-ph 9406356,
in {\it Electroweak Physics and the Early Universe},
eds. J. Romao and F. Freire, proceedings of Sintra conference,
March 1994, published by Plenum Press, New York (1994). }.
This has also been pointed out in a recent paper by CKN
\ref\cknsbnew{A. Cohen, D. Kaplan and A. Nelson,
Phys. Lett. {\bf B 336 } (1994) 41. }.
One of the objectives of the present  work
is to bring these previously unconnected pieces into one coherent
framework  which includes all the important effects and clarifies their
relation to one another.

This paper is organized as follows. Section 2 discusses the
WKB treatment of particle dynamics described by the Lagrangian
\lagrangian. In section 3 we introduce the Boltzmann equation
and discuss the fluid approximation with which we truncate it
to an analytically tractable form.  In the following  section we derive
the resulting fluid equations.  In section 5 we analyse
these equations in several steps, illustrating how perturbations
may be generated in front of the wall  and identifying  the parameters which
determine the behavior of the solutions. We derive the
reduced equations needed for
the calculation of baryon production in much of the
favored parameter space of wall thicknesses and velocities.
In section  6
we analyze these equations, treating the two
source terms separately, and calculate the resulting
baryon asymmetry in each case.  In section 7 we compare
the results we have found to those obtained in the thin
wall case in \JPTlonga. In section 8 we conclude with a
summary of the paper and a discussion of its shortcomings
and directions for future work.

\medskip

\centerline{\bf 2. WKB Dynamics}

The WKB approximation to the dynamics of particles in
the background of the bubble wall is good provided
the length scale on which this background varies is long  in
comparison to the de Broglie wavelength of the typical thermal
particles we wish to describe.
This is simply the requirement that the thickness of
the bubble walls  $L$ be greater than $T^{-1}$.
As we have
indicated above, this is a very reasonable expectation.

To describe the WKB `particles' we turn to the Dirac equation
derived from \lagrangian. The dispersion relation is obtained
as follows. In the rest frame of the bubble wall we assume that the
field $\tilde{Z}_\mu = (0,0,0,Z(z))$, and we can boost to a frame
in which the momentum perpendicular to $z$ is zero. In this frame
the Dirac equation reads (after multiplying through by $\gamma^0$)
\eqn\diraceqna{i\partial_0 \psi =
\gamma^0( -i\gamma^3\partial_z +m)
\psi
 - g_A Z \Sigma^3 \psi
}
where $\Sigma^3=\gamma^0 \gamma^3$ is the spin operator.
Setting $\psi \sim e^{-iEt +ip_z z}$, we see that the energy
is given by  the usual expression for a massive fermion plus a
spin dependent correction. The eigen-spinors are just
the usual free Dirac spinors.
Returning to the $p_\perp \neq 0$ frame amounts to replacing
$E$ with  $\sqrt{E^2-p_\perp^2}$, from which we find the
general dispersion relation
in the wall frame,
\eqn\dispersiona{
 E= [p_\perp^2 + (\sqrt{p_z^2 + m^2} \mp g_A Z)^2]^{1\over2} \quad
\Sigma^3=\pm 1
 }
where $\Sigma^3$ is proportional to the spin $S_z$ as measured in the
frame where
$p_\perp$  vanishes. The same dispersion relation
holds for antiparticles.
The particles we are most interested
in for baryogenesis are left handed particles (e.g. $t_L$) and right
handed
antiparticles ($\overline{t_L}$), since these couple to
the chiral anomaly. Note that they couple {\it oppositely}
to the $Z$ field.

In Figure 1  these dispersion relations are plotted for $p_\perp=0$
for (i) $m > g_A Z$, (ii) $m <  g_A Z$ and (iii) $m=0$. We see how the
branches deform into one another as we turn on the mass. In particular
we note how the left- (L) and right-handed (R) branches break up
into two pieces and form the $\Sigma^3 = \pm 1$ branches, as
the mass couples the two
chiral components on the wall. Correspondingly it is straightforward
to see how
the eigen-spinors (the usual Dirac massive eigen-spinors)
become chirality ($\gamma^5$) eigenstates
for $p_z >> m$. The conservation of spin on the wall
gives a simple picture of how this happens - an
in-going left-handed particle incident on the wall
evolves in an eigenstate of spin on the wall. If its momentum is
reversed it emerges as a right handed particle  since its spin
is conserved.

For what comes below we will find it useful to keep this identification
in mind. We label the states on the wall by the states they
deform into as the mass is turned off and write the dispersion relation
as
\eqn\dispersion{\eqalign{
     E^{L,R} &= [p_\perp^2 + (\sqrt{p_z^2 + m^2}
\pm sign (p_z) g_A Z)^2]^{1\over2} \cr
     &\rightarrow [p_\perp^2 + (p_z \pm g_A Z)^2]^{1\over2} \quad {\rm
as}
\quad m \rightarrow 0 \cr }}
where the $+(-)$ singles out the states which become $L\; (R)$ in the
unbroken
phase. (We assume the wall propagates from left to right, so that
incident particles from the unbroken phase initially have $p_z<0$.)
The anti-particle $\bar L$ of a left-handed particle $L$
is right-handed  and so we write the dispersion
relation for anti-particles
as
\eqn\dispersionap{\eqalign{
     E^{\bar L, \bar R}\equiv p_o^{\bar L, \bar R} =
[p_\perp^2 + (\sqrt{p_z^2 + m^2} \mp sign (p_z) g_A Z)^2]^{1\over2}. }}
In the WKB approximation we take each  particle to be a wavepacket
labelled by canonical energy and momentum $(E=p^0, \vec{p})$ and
position
$\vec{x}$. To compute the trajectory of such a wavepacket, we
first calculate the group velocity
\eqn\gpvel{v_i= \dot x_i= \partial_{p_i} E }
and second, using conservation of energy $\dot{E} =
\dot x_i \partial_i E + \dot p_i \partial_{p_i} E =0$, we find
\eqn\heqns{ \dot p_i = - \partial_i E}
Together these constitute Hamiltons equations for the particle.
The {\it momentum} of the particle is not a gauge invariant
quantity, but the particle worldline certainly is, and
we can for example calculate the acceleration from the Hamilton
equations.
We find
\eqn\acc{
{d v_z \over dt} = -{1\over 2} {\partial_z (m^2)\over E^2} \pm{\partial_z(g_A Z
m^2)
\over E^2 \sqrt{(E^2-p_\perp^2)} }  +o(Z^2)
}
%^M
where of course $E$ and $p_\perp$ are constants of motion. The first
term
describes the effect of the force due to the particle mass turning on,
the second the chiral force. In the massless limit the latter
vanishes, as it should because in this case the chiral gauge field
can just be gauged away.

We shall later need a few more explicit expressions, namely
\eqn\eom{\eqalign{ v_\perp= {p_\perp \over E} \qquad
v_z = {1\over E}\bigl [p_z \pm g_A Z {|p_z| \over \sqrt{p_z^2
+m^2}}\bigr ]\cr
\dot p_\perp = 0 \qquad \dot p_z=
{\sqrt{p_z^2+m^2} \pm sign(p_z) g_A  Z \over E}
      \bigl[ \pm sign(p_z) g_A \partial_z Z  +  {\partial_z m^2 \over
2\sqrt{p_z^2+m^2}}\bigr]. \cr }}
As $m \rightarrow 0$ we recover the equations of motion for the L and R
particles in the pure gauge field $Z$.

Before we move on let us remark on our neglect of
one effect in deriving these dispersion relations -
thermal corrections to the fermion self-energies.
In Appendix  C  we discuss how the dispersion relations
and consequently the force terms are
modified when one includes this effect. Depending
on  whether the $Z$ field is `strong' or `weak' in the
sense defined in the Appendix,
one can show that the
resulting dispersion relations are either those given here plus
small corrections, or slightly modified ones which lead to
force terms with altered momentum dependence. This is a
complication we will ignore in the rest of this paper.

\bigskip

\centerline { \bf 3. The Boltzmann Equation in the Fluid Limit}

Since most particles in the plasma are well described
by the WKB approximation with respect to their interactions
with the wall, we proceed to treat the fluid of such excitations
as consisting of
classical particles with definite canonical position and momenta,
and energy given by the derived dispersion relations. The
Boltzmann equation for the phase space density $f(\vec{p},\vec{x},t)$
is
\eqn\beqn{d_tf=\partial_tf+\dot{\vec{x}}\cdot \partial_{\vec{x}}f
  +\dot{\vec{p}}\cdot\partial_{\vec{p}}f= -C [f].}
The collision integral $C[f]$
describes how
the phase space densities are changed by interactions. The dominant
interactions  which we will consider are,
because of Debye screening, short ranged, and to a first approximation
may be treated as pointlike.
They are to be calculated at a given spatial point $\vec{x}$ using the
Dirac spinors appropriate to the local value of the
background fields $m(\vec{x})$, and $Z(\vec{x})$, taken to be constant.
The Boltzmann equation is in principle solvable, but
in order to make it analytically tractable we shall
we consider an approximation (truncation)  which we expect to
be quantitatively reasonable
based on a perfect fluid form for the
phase space density.
If interaction rates are fast, the collision
integral forces the phase space density towards a form
which minimizes it, namely local thermal equilibrium. For the case
of a single fluid this means the form
\eqn\fluida{f(\vec{p},\vec{x},t)=
{1 \over e^{\beta(\gamma(E -\vec{v}. \vec{p})) -\mu)}
\pm1 }}
where $\beta$, $\mu$ and $\vec{v}$ are functions of $\vec{x}$ and $t$,
and $\gamma=1/(1-v^2)^{1/2}$.
Imposing this form on the
left hand side of the Boltzmann equation and integrating to
find the moments, one arrives at a set of coupled
equations for the functions $\beta(\vec{x},t), \mu(\vec{x},t),
\vec{v}(\vec{x},t)$
which parametrize the phase space density. Expressed in terms
of the energy density  $\rho$, number density perturbation
$n$, pressure $p$ and velocity $\vec{v}$ these become the familiar fluid
equations
in the case of a free fluid:
\eqn\fluidem{\eqalign{\partial_t n + \vec{\nabla}\cdot [\gamma n\vec{v}] = 0
\cr
        \partial_t \rho  + \vec{\nabla}\cdot\bigl [(p+\rho)\gamma\vec{v}\bigr
] = 0 \cr
     \partial_t \bigl [ (p +\rho) \gamma\vec{v})\bigr ] +  \vec{\nabla}p= 0.
\cr }}
We are going to treat the approximately left-handed
excitations $L$  and their antiparticles $\overline{L}$
as two fluids, making an ansatz
 of the form \fluida\ for each.

This requires some justification, because
the dominant interactions which bring our
fluids to the form \fluida\
 are the same ones which damp
away the temperature and velocity differences
with respect to the background. This is not
true of the chemical potentials which are
only attenuated by slower chirality changing processes.
We keep the temperature fluctuation $\delta T$
in order to quantify the rate at which the form
\fluida\ is approached. We use the velocity
perturbation  in the form \fluida\ to model the
anisotropic response to the force.
We will in fact see that the  precise form
of this perturbation  does not enter the
final result, and its
only role is to allow the particles to move in response
to the force, and set up a chemical potential perturbation.

In the present case we wish to determine how different species are perturbed,
by source terms which will enter these equations in a way we will calculate.
In particular we must distinguish between particles and anti-particles as
it is the difference in the perturbations to these which is needed to source
baryon production through sphaleron processes. To do this we will
treat each particle species as a fluid described by the fluid ansatz
\fluida,
(in which the functions $\beta,\mu,\vec{v}$ are allowed to be different
for each species) which makes the self-interaction collision terms zero.
We then superimpose on  this the interactions of these fluids, which
will lead to terms in the equations damping all these perturbations
to the local thermal equilibrium for the whole fluid.

Is this a good approximation?
There are three conditions which must
be fulfilled:

(i)  The interactions must be fast enough to ensure the
system is maintained in the approximate form \fluida\ as the wall
moves. This should be a good approximation
if the time the wall takes to pass
is long in comparison to the time scale $\tau$ for the system to
attenuate fluctuations away from this form {\it i.e.\/} we want
${L \over v_w} > \tau$.
What should we take this time scale to be? We will see in
due course that this question
receives an answer within our calculation:
We should take the scale $\tau$ to be approximately $D$, the
diffusion length.

(ii) The rate at which the system is brought to this local
equilibrium form should
be faster than the rate at which the perturbations we keep in
 our ansatz
are attenuated. This is {\it not} the case, because
the same gluon exchange
processes which force $f$ to the form \fluida\
also damp away the velocity and temperature
perturbations. Thus \fluida\ is unlikely to be a very accurate
description of the precise form of those perturbations.
The chemical potential fluctuation will be the crucial term
in our answer, which shall determine the final baryon asymmetry,
and this is
attenuated only by processes which change the number of particles
in our `fluids'. As we will discuss at length these are
indeed typically much slower than the gauge boson exchange processes
which damp $\delta T$ and $\vec{v}$.
The temperature perturbation shall actually play little role
- we keep it merely to see
how thermalization occurs and to estimate its rate.

(iii) The mean free time for particle interactions should be
long compared to their inverse energies, in order that we can descibe the
physics in terms of a set of `free particle'
eigenstates of well
defined energy.  This condition means the `width' of a state $\Delta E$
should be much smaller than its energy $E$. The particles of
central interest in this paper have energies
$E \geq T$, and $\Delta E \sim g^2 T $,
so the condition is reasonably well satisfied even when one includes
the strong interactions.

 We should also mention a subtlety
about the distribution
\fluida. The form
is dictated by the quantities which are conserved in the
local interactions. That these are $E=p_o$ and $\vec{p}$
is a non-trivial fact in this case in which the
gauge symmetry is broken.
In the presence of a pure gauge field,
it is not immediately obvious whether one should take just the
canonical momentum, or perhaps some linear combination
with the gauge field, as the correct conserved quantity.

For a conserved gauge charge, which one takes is immaterial
since the difference can always be absorbed in the chemical potential,
which is determined by the condition on the conserved charge.
But here the gauge symmetry has been broken, and we
thus need to determine
what the correct conserved energy and momentum are in all
local
interactions, including those which violate the broken gauge charge,
which we assume to take place in approximately
constant background fields. By a simple calculation  one can show
that the `canonical' stress tensor
\eqn\emtensora{\Theta^{\mu\nu}=\bar\psi\gamma^\mu i \partial^\nu\psi}
is in fact conserved in constant background fields, since
from the Dirac equation derived from \lagrangian,
\eqn\emtensorb{\partial _\mu \Theta^{\mu\nu}=(\partial^\nu  m )\bar\psi
\psi
 + g_A (\partial_\mu \tilde{Z}^\nu )\overline{\psi}\gamma^\mu \gamma^5
\psi
}
The conserved energy and momentum are then easily calculated from
the eigenstates we discussed in the previous section, and
are indeed just the canonical energy $p^0=E$, and momentum $\vec{p}$.
This is as to be expected from translation invariance.

We also need to specify what precisely the fluids are as this
is ambiguous once we turn on the background in which the dispersion
relations and particle eigenstates become different. In the unbroken
phase
we take the fluids to be the chiral eigenstates. Which states
described  by the dispersion relations in Figure 1 do we take to make
up the fluids described by the phase space density in \fluida ?
The answer is that foreshadowed by the way we wrote the
dispersion relations in \dispersion. The motivation for this
choice is that most particles in these fluids are then to a good
approximation in a given chiral state. This of course breaks
down completely for low momentum states $p_z \sim m$ which are
effectively equal mixtures of the chirality eigenstates.
By dividing the fluids in this way we also misdescribe the dynamics of
these
low momentum states by connecting the wrong branches at $p_z=0$.
Most particles in the fluid will not change direction between
scattering and so this should be a small effect. We will return
to these points later. The essential point is that the effects
we describe will be dominated by particles at thermal energies.
The force is felt by all particles in the plasma, not just those
at low momentum.

\medskip

\centerline{ \bf  4. Fluid Equations}

We now proceed to derive in detail the truncation of the Boltzmann
equations
which we have just described. The substitution of the ansatz \fluida\
in the left hand side of \beqn\ gives, in the rest frame of the wall in
which the energy is time independent,
\eqn\feqnla{- \bigl [\beta \partial_t \mu + \beta p_z \partial_t v
-\partial_t \beta (E-\mu-vp_z)\bigr ]f'
            - \bigl [\beta \partial_z \mu + \beta p_z \partial_z v
-\partial_z \beta (E-\mu-vp_z)\bigr ]f' \partial_{p_z} E + \beta v
f'\partial_z E }
where $f'={d\over dx}({1\over e^x \pm 1})$ and  $x=\beta(E-\mu-vp_z)$.
 We take $\vec{v}=v \vec{z}$ and work to leading order in $v$.
Making the substitution $\bar\mu=\mu \pm v g_A Z$ (with the sign chosen
appropriately for the left-handed and right-handed
particle and antiparticle fluids ) and writing $p_z\pm g_A Z=k_z$,
the physical momentum
in the zero mass limit,  we then have
\eqn\feqnlb{\eqalign{ - \bigl [\beta \partial_t \bar \mu + \beta k_z
\partial_t v -\partial_t \beta (E-\bar\mu-vk_z)\bigr ]f'
            - \bigl [\beta \partial_z \bar \mu + &\beta k_z \partial_z v
-\partial_z \beta (E-\bar \mu-vk_z)\bigr ]f' \partial_{p_z} E \cr
      +& \beta v f'\bigl[\partial_z E \mp g_A \partial_z
Z.\partial_{p_z}E \bigr].\cr }}
We make this change in variables so that our equations will be in the
familiar form \fluidem\ as $m \rightarrow 0$. The variable $\bar \mu$
is really what one would usually call the chemical potential in the
$m=0$ limit, as
fluctuations in number density are proportional to it. The $\mu$
in \fluida\ is in fact a gauge dependent object and $\bar \mu$
is simply the appropriate gauge invariant chemical potential
which appears in real physical quantities calculated from
the distribution function.

The crucial term here
is the coefficient of the velocity $v$. We are working in the rest
frame of the wall and so write $v=-v_w + \bar v$ since we are
interested
in the perturbations to the background which is a plasma moving by the
wall
at velocity $-v_w$ in this frame. Then we see that there is
a source term proportional to $v_w$ with coefficient
\eqn\forcea{\eqalign{ \beta
      {\sqrt{p_z^2+m^2} \pm sign(p_z) g_A  Z \over E}
      \bigl[ \pm sign(p_z) g_A \partial_z Z ( 1- {|p_z| \over
\sqrt{p_z^2+m^2}} ) +  {m\partial_z m \over \sqrt{p_z^2+m^2}}\bigr]
}}
 where
$\pm$ correspond to the signs in the dispersion
relations \dispersion\ and \dispersionap\ for
the particles and anti-particles in
our (approximately) chiral fluids.
This source manifestly vanishes as $m \rightarrow 0$ and when integrated will
give us the force on the plasma when the wall is pushed through it.
Before doing this integration of these equations we consider also
the right hand side of the Boltzmann equation.

The collision integral $C[f]$ in
\beqn\ is equal to the rate of change of the
phase space density $f(\vec{p},{x})$ due to collisions.
Considering only processes  in which there are two
incoming (labelled
as $1$ and $2$) and two outgoing particles (labeled as
$1^\prime$ and
$2^\prime$) we can write it in the following form
\eqn\collint{\eqalign{ \sum_{processes}
    {1\over 2 (E_1 - v_w p_z)}\int_{p_{1^\prime},p_2,p_{2^\prime}} |{\cal M}|^2
 (2\pi)^4 {\delta}^4 \Bigl (\Sigma_i{ p_i}\Bigr )  {\cal P}[f_i]\cr
{\cal P}[f_i]=
\left [ f_1 f_2 (1\mp f_{1^\prime})(1\mp f_{2^\prime})-
f_{1^\prime} f_{2^\prime} (1\mp f_1)(1\mp f_2)\right ] }}
where the precise spinors used as eigenstates will determine the
various
normalizations in the integral, $ |{\cal M}|^2 $ is the matrix
element for the process, and $\mp$ is for fermions (bosons).
Taking the fluid ansatz \fluida\ for each particle species and doing a
perturbative expansion in  $\beta={1\over T}$ (about $\beta_o={1\over
T_o}$), ${\mu \over T_o}$ and $v$ we get, to leading order,
\eqn\collintb{\eqalign{\approx \sum_{processes}&{1\over 2 (E_1 - v_w p_z)}
    \int_{p_{1^\prime},p_2,p_{2^\prime}}  |{\cal M}|^2
 (2\pi)^4 {\delta}^4 \Bigl (\Sigma_{ p_i}\Bigr )
    \left [ f_1 f_2 (1\mp f_{1^\prime})(1\mp f_{2^\prime})\right ] \cr
      & \times \beta_0  \left [{(\delta T_1-\delta T_2)\over T_0}
(E_1-E_{1^\prime})
            +(\vec v_1-\vec v_2)\cdot (\vec p_1-\vec p_{1^\prime})
       +\sum (\mu_i)\right ]\cr }}
where $f_1$ {\it etc.\/} are now the unperturbed distribution functions at
temperature $T_o$. We have performed this expansion in the
{\it plasma} frame variables for reasons we will explain below when we come
to integrate this expression. The normalization factor is just the
energy in the wall frame.
$\sum (\mu_i)$ means a sum over chemical
potentials with a positive (negative) sign for in-going
(outgoing) states.
In the temperature and velocity terms we have assumed that
the in-going and out-going $1$ and $1'$ are in the same fluid and
the same of $2$ and $2'$,  since this is the case for the dominant
scattering processes.
For each term in brackets we must take
the fastest process which force these fluids to the same
thermal equilibrium. The ones that attenuate the
temperature and velocity perturbations are gluon exchange diagrams
(for quarks) or weak boson exchange (for leptons), shown in
Figure 2.
These processes do not contribute to the chemical potential damping
however since they do not change particle number. Examples
of processes
which do contribute are
the helicity flipping gluon exchange process (as in Figure 2,
but in the presence of a mass term)
which can occur on the wall,
and the Higgs mediated decay like that in Figure 3, as well
as sphaleron processes (both strong and weak).

We now integrate the Boltzmann equation over
$\int {d\!\!\!^-}^3p$, $\int  {d\!\!\!^-}^3p E$
and $\int  {d\!\!\!^-}^3p \, p_z$
( {\it wall} frame variables) to get three moments which
give us three first order differential equations for the
three functions ${\delta T \over T_o}, \bar v, { \bar \mu \over T_o}$
characterizing the perturbations
in each fluid.  In Appendix A the terms obtained by
integrating the collision integral are
analyzed. Considerable simplifications occur
provided the integrations are done in the plasma frame
variables in which $f_o$ has the standard (unboosted) form.
Thus we take the linear combinations
$\int  {d\!\!\!^-}^3p (E + v_w p_z)$
and $\int  {d\!\!\!^-}^3p \, (p_z +v_w E) $ of the latter two
integrations which are integrations over the energy
and momentum in the plasma frame (to leading order in $v_w$).
The velocities which appears in \collintb\
are those in the plasma frame, but the difference
of velocities in this frame is equal to the difference
of the velocities in the wall frame to leading order in $v_w$.
 We work perturbatively in
${\delta T\over T_o}, \bar  v, { \bar \mu \over T_o}$
keeping the leading terms in these quantities and their derivatives,
dropping next order corrections to these coefficients in
${Z \over T}$ and ${m \over T}$.

The equations which result are
\eqn\fluideqns{\eqalign{
-v_w ({\delta T\over T_o})^\prime -a v_w
({\bar\mu\over T_o})^\prime+{1 \over 3} \bar v^\prime + g v_w({m^2\over
T_o^2})'  & =
-\bar \Gamma_ \mu \sum {\bar \mu\over T_o}
-\Gamma_\mu \sum
({\bar \mu\over T_o} \pm v_w g_A {Z \over T_o}) \cr
- v_w({\delta T\over T_o})^\prime
-b v_w ({\bar\mu\over T_o})^\prime +{1 \over 3} \bar
v^\prime
+h v_w({m^2 \over T_o^2})' &= \cr
 -\Gamma_T\sum {\delta T\over T_o} -\bar \Gamma'_\mu
\sum {\bar \mu \over T_o} - &\Gamma'_\mu \sum
({\bar \mu\over T_o} \pm v_w g_A {Z \over T_o}) \cr
 ({\delta T\over T_o})^\prime +
b  ({\bar\mu\over T_o})^\prime - v_w \bar v^\prime
  \pm c  v_w {(g_A Z m^2)^\prime \over T_0^3}
        &= -\Gamma_v  \sum \bar v \cr }}
where $a={2\zeta_2 \over 9\zeta_3}$,
$b={3\zeta_3 \over 14\zeta_4}$,
$c={\ln 2 \over 14\zeta_4}$, $g={\ln 2 \over 9 \zeta_3}$,
$h={\zeta_2 \over 42 \zeta_4}$ and $\zeta_n$ are the Riemann
$\zeta$-functions ($\zeta_2={\pi^2 \over 6}$, $\zeta_3=1.202$,
$\zeta_4=\pi^4/90$). To derive  these equations
in this form we have  expanded  $E,\partial_zE,\partial_{p_z}E$
 around $m=0$. Except the force terms
all the terms on the LHS have just the free fluid coefficients -
precisely what the equations \fluidem\ give if one expresses them in
terms
of $\delta T, \mu, v$ ( for $m=0$ i.e. $p={1 \over 3}\rho$).
The positive sign in the force term applies
to the $L$ and $\bar R$ fluids, the negative sign
to the $R$ and $\bar L$ fluids; the mass is that of the
appropriate fermion.

The terms on the RHS require
some explanation. We have written
\eqn\collintexp{\eqalign{
{1 \over 3n_0}\int d{\!\!^-}^3p C[f])  & =
\bar \Gamma_ \mu \sum {\bar \mu \over T_o}
+\Gamma_\mu \sum ({\bar \mu\over T_o} \pm v_w g_A {Z \over T_o})  \cr
{1\over 4\rho_0}\int d{\!\!^-}^3p p_o C[f] &=
\Gamma_T\sum {\delta T\over T_o}+ \bar \Gamma'_ \mu \sum
{\bar \mu \over T_o}
+\Gamma'_\mu \sum({\bar \mu\over T_o} \pm v_w g_A {Z \over T_o})  \cr
{3\over 4\rho_o}\int d{\!\!^-} ^3p p_z C[f] &=\Gamma_v  \sum \bar v
\cr }}
where $n_o$ and $\rho_o$ are the unperturbed number and energy
densities respectively at $T_o$.

To arrive at the RHS  of these equations
we have used symmetry arguments to show
that various terms are zero for tree level processes we are interested in.
This is discussed in Appendices A and B.

In the collision terms
as written in \collintb\ we see that the substitution of the variable
$\bar v$ gives the same expression with $v$ replaced by $\bar v$ as
it is only the relative velocity which is damped by this term. However
the substitution of $\bar \mu = \mu \mp v_w g_A Z \pm \bar v g_A Z$
is very non-trivial. The latter piece gives only a higher order
correction to our equations. But the term $v_w g_A Z$ only drops out
to give the same expression with $\mu$ replaced by $\bar \mu$
if the charges $g_A$ on  the external legs in the process
sum to zero. In the limit that the vevs of the
Higgs fields vanish
$Z$ is a pure gauge field for the fermions and Higgs particles,
and so this cancellation occurs for all
processes which conserve any linear combination
of electric charge and hypercharge.
Thus $\bar\Gamma _\mu$ and $\bar\Gamma' _\mu$ are calculated from
decay processes
which conserve hypercharge, $\Gamma _\mu$ and $\Gamma' _\mu$ from
processes which explicitly violate it, the latter picking up
a net contribution proportional to $v_wZ$.
These latter rates, being hypercharge (and $T_3$)\footnote{$^\dagger$}{The
fact that we refer to these processes as
`hypercharge violating' has no particular significance.
They violate any linear combination of hypercharge and
electric charge. In particular they violate the axial charge
$g_A$ which gives the coupling of the CP violating condensate
to the fermions and Higgs fields.}
violating,
are vev squared suppressed and therefore  any baryon asymmetry
produced by them will vanish as the vevs do.
When the vevs are nonzero, the  $Z$ field perturbs
hypercharge violating processes out of equilibrium locally,
while the hypercharge conserving processes remain in equilibrium,
simply because they conserve the charge associated with the
gauge field. These latter ``see''  $Z$ as a pure gauge mode,
not a real gauge invariant field  which shifts the energies.
This is precisely the type of effect which CKN
called `spontaneous' baryogenesis. We will discuss in section 6.2
what light this treatment throws on the question raised
by Dine and Thomas  in \dt\ about how the background
should be modeled.

Before we move on to solve \fluideqns\  for some specific cases
we make a few general comments:

(i) We have dropped all the time derivatives because we are interested
in
stationary solutions. If we wish to
understand, for example,  how the stationary solutions are set up
at the time of nucleation, we include the terms which are
obtained from the time derivatives in
\feqnlb\ after integration.

(ii) We have dropped all long range fields which result from the
perturbations. This amounts to neglecting the effect of screening
of electric charge and hypercharge on the solutions we will study.

These terms can easily
 be included (through an addition to $\dot{\vec{p}}\;$) and give a
 term proportional to the field $\vec{E}$  in
 the third equation. Making use of Gauss' Law this
can be written in terms of the perturbations,  thus coupling
the LHS of all the equations to one another (see
 \JPTlonga\
section 7). We will comment further
on this point  in the conclusions.

(iii) As $v_w \rightarrow 0$ the only solution unperturbed at infinity
is the trivial solution $\beta=\beta_o$ and $v=\mu=0$. Any perturbation
which may source baryon number (or indeed push any
process out of equilibrium) arises due to the motion of the wall.
This does not mean however that this static equilibrium is the
unperturbed one which pertains when the field is not present. This
is because the energy in the distribution function is the perturbed
energy.
For example if we calculate the number density
$\int {d\!\!\!^-}^3pf|_{\mu=v=0}$ and subtract the true unperturbed
number density we do not get zero if we integrate e.g. over states
 with $p_z > 0$.
This equilibrium has in it an excess of right-moving
spin ${1\over 2}$ particles and an equal and opposite underdensity
of left-moving spin $-{1 \over 2}$ particles. The way to understand
this
is by analogy with an electromagnetic potential which is
screened.
The thermal equilibrium reached in the presence of such a potential has
an overdensity of particles in proportion to their charge. Here the
``screening'' of the force induced by turning on this field has
the effect of dragging in particles as described by this distribution
function with $v=\mu=0$. This was the essential point made in \JPTtran,
except that the energy perturbation was modeled (inappropriately)
by a purely fermionic hypercharge potential.

(iv) As the vevs of the Higgs fields vanish the only solution is again
the trivial one (without time dependence)
since both the mass and the rates of hypercharge violating
processes go to zero.  In this case the distribution functions do also
describe the true unperturbed plasma since in this limit
the dispersion relation  approaches the pure gauge one in \dispersion.

 (v) We will explain below that in the case that we neglect the
temperature fluctuations the system is reduced to the first and
third equation in \fluideqns. Dropping all the force terms and
setting all the decay rates to zero, to order $v_w$ we
obtain
\eqn\diffusion{ b  ({\bar\mu\over T_o})''
    = -3a v_w \Gamma_v  ({\bar\mu\over T_o})' }
which describes pure diffusion $\dot n = - D \nabla^2 n$.
We can then read off the relation $D={b \over 3a}\Gamma_v^{-1}$.

\medskip

\centerline {\bf 5. Solutions of Fluid Equations}

We now turn to the analysis of the fluid  equations \fluideqns,
with  the goal of  understanding the baryogenesis which results
from the perturbations they describe.
As discussed in section 2 of the accompanying paper
\JPTlonga\ the  anomalous baryon number violating
process is  perturbed from equilibrium  by
a difference in the distribution functions of
left-handed fermions and their (right-handed) antiparticles.
As we wish to study baryon production we
therefore take  the difference
of \fluideqns\ for left-handed particles
and their anti-particles, and get
\eqn\feqns{\eqalign{
-v_w ({\delta T\over T_o})^\prime -a v_w
({\bar\mu\over T_o})^\prime+{1 \over 3} \bar v^\prime  & =
-\bar \Gamma_\mu \sum ({\bar \mu \over T_o})-\Gamma_\mu \sum (v_w g_A
{Z \over T_o}) \cr
-v_w({\delta T\over T_o})'
-b v_w({\bar\mu\over T_o})^\prime +{1 \over 3} \bar v'
&= -\Gamma_T({\delta T\over T_o}) -\bar \Gamma'_\mu \sum ( {\bar \mu
\over T_o})
-\Gamma'_\mu \sum (v_w g_A {Z \over T_o})\cr
 ({\delta T\over T_o})^\prime +
 b  ({\bar\mu\over T_o})^\prime - v_w \bar v'
  +F(z)  &= -\Gamma_v  \bar v \cr }}
We have  compactified our notation:
$\bar \Gamma_\mu$ now includes {\it all} decay processes
and $\Gamma_\mu$ denotes only the hypercharge violating ones.
The force terms in the first two equations  in \fluideqns\
cancel out because the gradient in the real mass affects particles
and anti-particles equally. The parameters
$\delta T, \bar v, \bar \mu$ now represent  the {\it difference}
in these quantities for particles and antiparticles, and the
force
\eqn\forceterm{F=2 cv_w g_A {( Z m^2)^\prime \over T_0^3}
\equiv Av_w {( Z m^2)^\prime \over T_0^3}
}
We have used the
fact that particles  and anti-particles couple in  exactly
the same way in the gauge boson exchange diagrams to cancel
out the  temperature and velocity perturbations of the other
fluids. This removes the sum in the
$\Gamma_T$ and $\Gamma_v$ terms. The counting factor which results -
over
particles and anti-particles of all flavors - has been
absorbed in the definition of $\Gamma_T$ and $\Gamma_v$.
For quarks we show in Appendix  A that
\eqn\rates{\eqalign{   \Gamma_v = 3\Gamma_T \simeq \alpha_s^2
{\ln}{1\over \alpha_s} T\simeq {T\over 20}
}}
We note that the relation  $D=b/3a\Gamma_v$
from \diffusion\ then gives $D \simeq 5/T$
in very good  agreement with the value calculated by a different
method in \JPTlonga.
The only coupling to perturbations in other fluids remains in the
sums for the decay processes.
There are two distinct sources in \feqns\ for the perturbations:

$\bullet$ The force terms $F(z)$.

$\bullet$ The hypercharge violating processes which are perturbed from
 equilibrium as the wall passes when $v_w Z \neq 0$.

As the equations are linear we can separate these sources and study
them
independently.

We will not attempt to solve the full set of equations
in complete generality for each of these two source terms.
Firstly, we will limit our scope by considering only
the case where the source terms directly affect the
{\it top quark}. Both the classical force term
and (we will see below) the spontaneous baryogenesis terms are
proportional to the mass squared of the fermion
for which \feqns\ describe the particle minus
anti-particle perturbations. We will thus work
in the approximation that only the top quark
Yukawa coupling is non-zero, which is good
if the fermion Yukawa couplings are (as in
the minimal standard model) proportional
to their zero temperature masses. We
will not consider the case
emphasized in \JPTlonga, where in a two Higgs
extension the tau lepton can have a Yukawa
coupling as large as that of the top quark
(and hence a comparable finite temperature
tree level mass).

We now discuss:

(i) Dropping the
temperature fluctuations, so reducing the equations
to just two coupled equations for $\bar \mu$ and $\bar v$,
which can in turn be written as two second order uncoupled
equations for the these two variables.

(ii) The solution of
the reduced equations for the classical force source
term with the simplification that we neglect all decay
processes.

(iii) The problem with decay processes
included, for the classical force source term.

(iv) The source terms from hypercharge violating processes
and how they source baryogenesis.

{\bf 5.1 Thermalization and Validity of the Fluid Ansatz}

In section 3 we explained that we do not expect the
temperature fluctuations to accurately reflect the
perturbations in the plasma, because the tree-level
gauge boson exchange processes which damp them away
are exactly the same processes which damp away the
perturbations to the distribution functions
which we neglected in taking the ansatz \fluida.
We kept them in our ansatz however in order to
have a quantitative measure of the validity of our
approximation. It is this point which we first consider here.

Neglecting all decay processes
the equations \feqns\ are

\eqn\feqnsnd{\eqalign{
-v_w {\delta T'\over T_o} -a v_w
{\bar\mu '\over T_o}+{1 \over 3} \bar v^\prime  & = 0
 \cr
-v_w{\delta T'\over T_o}
-bv_w{\bar\mu '\over T_o} +{1 \over 3} \bar v'
&= -\Gamma_T {\delta T\over T_o} \cr
 {\delta T '\over T_o} +
 b  {\bar\mu '\over T_o} -  v_w \bar v'
  +F(z)  &= -\Gamma_v  \bar v \,.}}
One can convert these to a pair of second order
uncoupled equations for $\delta T$ and $\bar \mu$
(by integrating the first equation directly to find $\bar v$),
and then solve them for various sources $F$.
We will not present here the details of this calculation
as the results are of no importance, except for
the following point. Two parameters - $\lambda_D L$ and
$\lambda_T L$ -  enter in determining the
behaviour of the solutions,  where
\eqn\roots{\lambda_D = { v_w \over D}  \qquad
\lambda_T =  { b  \over b - a }{\Gamma_T \over v_w}  }
 are  the two  roots of \feqnsnd\
and $L$ is the thickness of the wall.
The diffusion root $\lambda_D$ describes the diffusion tail in front of
the wall, while $\lambda_T$ describes the decay of perturbations behind the
source.
$\lambda_D L$ is simply the squared ratio
of the wall thickness to the distance a particle with
diffusion constant $D$ diffuses as the wall passes
($\sim \sqrt{Dt} \sim \sqrt{DL/v_w}$); this is the
parameter which, as we discuss below, will characterize
what we call `good transport' or `poor transport'.
$\lambda_T L$ is the ratio of the time of passage of
the wall ($L/v_w$) to the mean free time for temperature
attenuating processes ($\sim \Gamma_T^{-1}$)
and it is a measure of how efficiently the temperature perturbations
are damped on the wall.
One finds that when $\lambda_T L > 1$ the temperature perturbations
in the solutions to \feqnsnd\ are damped by at least
${1 \over \lambda_T L}$ relative to the chemical
potential fluctuations. Examining \feqnsnd\ one can in fact
see this damping directly in the second equation.
Taking all the derivatives to go as $L^{-1}$, we see
immediately that the suppression follows.
When one incorporates the decay processes
a similar conclusion follows provided
$\Gamma_T >> \Gamma_\mu$, which applies
(see rates given below).

Because this condition for the damping of temperature
fluctuations is the same as that of the validity of
our initial ansatz we always take it to apply and
reduce our equations to the first and third in \feqns\
with $\delta T$ set equal to zero. Noting the relation
between $\Gamma_T$ and $\Gamma_v$ from \rates\ and
$\Gamma_v={b \over 3a}D^{-1}$ from \diffusion, the condition becomes
\eqn\condtherm{ v_w < {L \over 3D} \qquad {\rm thermalization.}}
Still neglecting the decay processes, the equations simplify
further:
\eqn\feqnsndnt{\eqalign{
 -a v_w
{\bar\mu '\over T_o}+{1 \over 3} \bar v^\prime  & = 0
 \cr
 b  {\bar\mu '\over T_o} - v_w \bar v'
  +A v_w {( Z m^2)^\prime \over T_0^3} &= -\Gamma_v  \bar v .\cr }}
where $A=2c g_A$.
The solution to the first equation gives $\bar v= 3a v_w  {\bar\mu\over
T_o}$
(using the boundary condition that $\bar \mu=\bar v=0$ far
in front of the wall) and substituting in the second equation,
again using the relation $\Gamma_v={b \over 3a}D^{-1}$ from \diffusion,
we get  (to leading order in $v_w$)
\eqn\purediff{D {\bar \mu\,{'}\over T_o}
+  v_w {\bar \mu  \over T_o} =
- {D \over b} F(z) }

As discussed in section 3 $\bar v$ is, like $\delta T$,
not in itself to be taken to accurately describe
the perturbations in the fluid. This is the case
because we would expect there to be other anisotropic (in momentum)
components of a general distribution function which we have
neglected in our ansatz, which
will be damped away by the same
(tree-level gluon exchange) processes at
approximately the same rate as the perturbation parametrized
by $\bar v$ which we have taken. We cannot however
consistently set it to zero in \feqns; but as we now see
$\bar v$ is indeed damped by
$v_w$ relative to $\bar \mu$, so that it itself will
not contribute (at leading order in $v_w$) to the
biasing of the baryon number violating processes,
and neither does it enter in \purediff\ which  determines
$\bar \mu$. Its only role is to mediate the
force to the chemical potential, and we assume that
any other anisotropic component would have led to
approximately the same result.

The fluid equations \feqnsnd\ are calculated to leading order
in $v_w$. A fuller analysis incorporating all orders in $v_w$
can be performed and shows that the velocity at which
the leading order analysis breaks down is the speed of sound
$v_w\sim v_s= 1/\sqrt 3$
in the plasma.
If the wall moves faster than this there is no
solution in front of the wall  and perturbations cannot
propagate into this region. We are interested
in the case when perturbations can propagate in front
of the wall where the anomalous electroweak processes
are unsupressed. Thus we assume
\eqn\condvs{ v_w < v_s= {1 \over \sqrt{3}}  \qquad {\rm }}

\vfill\eject
{\bf 5.2 Classical Force sourced Perturbations without Decay}

We now solve \purediff\  for
a chosen ansatz for the force term in order to illustrate
the behaviour of the perturbations they describe and to gain
some simple intuition for the more complex case
where we include the  decay terms.
We take the following `ramp' ansatz
for the source term:
\eqn\ansatza{ Av_w {Z m^2\over T_o^3}= \cases { F_o (z-{L\over 2})  &
          $-{L\over 2} < z < {L\over 2}$  \cr
  0 & \rm {otherwise} \cr }     }
where $F_o$ is the constant force on the wall, and as above $A=2 c
g_A$.
\purediff\ has then a particular solution which is
non-zero only on the wall, and a homogeneous solution which
is an exponential describing  pure diffusion
$\sim e^{- {v_w  \over D}z}$. We impose
the boundary conditions
\eqn\bcsforce{\eqalign{
D ({\bar \mu \over T_o})
 \big | ^{{L \over 2} + \epsilon }_{{L \over 2} - \epsilon }= 0  \qquad
D ({\bar \mu \over T_o}) \big | ^{-{L \over 2} + \epsilon }_{-{L \over
2} - \epsilon }= {DL F_o \over b}
    } }
obtained by integrating \purediff\ taking $\bar \mu$ to be
at most step discontinuous across the boundaries
(so that its integral is continuous).
Solving, requiring
$\bar \mu$ to be finite at $- \infty$, we get
\eqn\solnscasea{\eqalign{
{ \mu  \over T_o} &= {F_o L \over b } \cases {0 & $z<-{L \over 2}$
\cr
 -{D \over v_w L}+ ({D \over v_w L}+1) e^{-{v_w \over D}(z+{L \over
2})}  & $-{ L \over 2}< z<{L \over 2}$ \cr
 [ e^{-{v_w \over D} L}
-{D \over v_w L}(1-e^{-{v_w \over D} L})] e^{-{v_w \over D}
(z- {L \over 2})}& $z>{L \over 2}$ \cr} } }
These solutions for ${\bar \mu \over T_o}$ are sketched in Figure 4
for the two cases (i) ${v_w L \over D} << 1$ and (ii) ${v_w L \over D}
>>1$,
when the solutions on the wall can be written
\eqn\limit{ {  -F_o L \over b } \cases
{ {1 \over 2}{v_w L \over D} + {( z - {L \over 2}) \over L}
\big [1- {1 \over 2}({v_w L \over D})^2 \big ] & ${v_w L \over D} << 1$
\cr
 { D \over  v_w L } & ${v_w L \over D} >> 1$ \cr }}
In the first case
we see that the solution mimics the
behaviour of the driving force. From
\feqnsndnt\ one can see that this behaviour
is generic for ${v_w L \over D} <<1$, for expanding in
$v_w$ one finds at leading order
the solution
\eqn\vlimit{  \bar v = 0 \,,\qquad {\bar \mu \over T_o}
= -v_w {A \over b}  {Z m^2\over T_o^3}  }
is non-zero only on the wall.
What the parameter ${v_w L / D}$ tells us is how
efficient the diffusion is in bringing us to \vlimit.
As discussed in section 5.1 it is simply
a ratio of the thickness of the wall to the distance
a particle diffuses as the wall passes. This  is
the parameter which defines `good transport'.
The corrections
to \vlimit\ describe a diffusion tail
in front of the wall which has amplitude
\eqn\ampla{{\bar \mu \over T_o}\big |_{L\over 2}
= -{1 \over 2}{v_w L \over D} \; { F_o L \over b} }
The factor of two arises from the fact that the
diffusion in front of the wall is driven by the average amplitude
of the potential on the wall, since particles are
in this regime ``seeing'' the whole wall.
This is confirmed (see examples later) by calculating
with other ansatzes.
We thus conclude that in more general
we would find the solution in front of the wall
\eqn\amplb{{\bar \mu \over T_o}
=  +{{v_w L \over D}}\; v_w {A\over b}  {\langle Z m^2\rangle \over
T_o^3}
e^{-{v_w \over D}(z - {L\over 2}) } }
where $\langle ..\rangle$ means the average value on the wall.

When   ${v_w L \over D} >>1$, i.e. when the transport is
`poor' over the relevant timescales,  the solution looks
like that in Figure 4 (ii).
The amplitude on the wall rapidly approaches that
in front of the wall which is
\eqn\amplc{{\bar \mu \over T_o}\big |_{L\over 2}
=  - {D \over v_w L } \; { F_o \over b} }
so that the integrated amplitude
$\int_{L \over 2}^\infty \bar \mu $ is suppressed relative to the
`good transport' case
by $({D \over v_w L })^2$.
It is this integrated amplitude of $\bar \mu$ (recall that
it is the {\it difference} of particle and anti-particle
chemical potentials) in front of the
wall which will be the effective driving force for the final
baryon asymmetry, when we assume the baryon number
violating processes are immediately switched off on the wall.
We note that the solutions \solnscasea\ have the
property that
\eqn\total{\int_{-\infty}^{+\infty} \bar \mu(z) dz= 0 }
In fact this can be shown directly by integrating \purediff\ once,
taking the solutions to be zero at $\pm \infty$.
The integral of the chemical
potential is zero because there is no net particle minus
antiparticle creation in the absence of the decay processes.
Thus the integrated contribution in front of the wall
exactly cancels that on the wall, the excess of particles
pulled onto the wall in response to the force being
exactly cancelled by a
deficit in the diffusion tail in front of the wall.
In terms of baryon production it follows that if
the sphaleron rate were unsuppressed
on the wall exact cancellation would occur between
the production in front of and on the wall
 so the source would not bias net baryon production. The extent of
the cancellation which can occur is  a sensitive function of the wall
profile, but unless $Z$ condenses only  where the vev
is very small this will not be significant. We will
always assume that the baryon number violating
processes are only turned on in front of the wall.

This property \total\ allows us to read off simply from
\vlimit\ the integrated amplitude in front of the wall
when ${v_w L \over D}<<1$ (to leading order in
this parameter) as
\eqn\resultnd{\int_{{L \over 2}}^\infty {\bar \mu \over T_o}
=  v_w {A \over b} \int_{wall} {Z m^2\over T_o^3} }
where again $A= 2c g_A$ ($c$ and $b$ being the numbers
defined after \fluideqns\ ). This in fact also
proves \amplb\ since we know that the solution
in front of the wall $\sim e^{-{v_w \over D}z}$.
We note that the result
is in this regime independent of $D$ and $L$, and
we will see that the factor of $v_w$ also drops out
when we calculate the baryon asymmetry.
We will defer doing so explicitly until after the
next section in which we discuss incorporating the
decay processes in this analysis.

{\bf 5.3 Decay Processes and Baryon Production}

Putting back the decay rates our equations are
\eqn\redfeqns{\eqalign{ -a v_w ( {\bar \mu \over T_o})^\prime
+{1 \over 3} \bar v^{\prime} &= -\bar \Gamma_ \mu \sum {\bar \mu \over
T_o}
+ \Gamma_\mu \sum v_w g_A {Z \over T_o}  \cr
b({\bar\mu \over T_o})^\prime  +  F(z) &= -\Gamma_v  \bar v  \cr  }}
where we follow again the conventions in \feqns. $\bar\mu$
is again the {\it difference} in particle and
anti-particle chemical potentials and the sums are
over the external legs of the processes which change
the number of particles minus anti-particles. We
have a pair of such equations for each fermion
of (approximate) chirality, where
$F= \pm Av_w {( Z m^2)^\prime \over T_0^3}$
($+$ for the $L-\bar L$ perturbations, $-$ for
the $R-\bar R$  perturbations) and $m$ is the
mass of the fermion, which we have assumed only
to be non-zero for the top quark.
The equations are coupled only through the sums in the
decay terms on the right hand side, and it is this coupling
which we now consider.

As before we can decouple the variables
$\bar \mu $ and $\bar v$ and concentrate
on a set of  second order equations for the $\bar \mu$:
\eqn\diffdecay{
D{\bar \mu ^{''}\over T_o}
+  v_w {\bar \mu  ^{'} \over T_o}
- \bar \Gamma_\mu \sum {\bar \mu  \over T_o} =
- {D \over b} F'(z)  -  \Gamma_\mu \sum  v_w g_A {Z \over T_o} }
The decay rates here have absorbed a factor of $1 /a$ relative
to the rates defined in \collintexp.
There are a similar set of equations for
the $\bar v$. We will not consider these
further since, as discussed
in the section 5.1,
$\bar v$ is attenuated relative to $\bar \mu$
and hence also its effect on the
baryon number violating processes
which is what concerns us here.

What are the decay processes? Recall that
our particle and anti-particle states are
those of our (approximately) chiral fluids.
They are the WKB eigenstates of the dispersion
relations  \dispersion\ and \dispersionap, which become pure
chirality states in the unbroken phase.

In the broken phase they are  helicity
states which are mixtures of the two chiralities.

It is thus natural that we divide the processes into
(i) those which occur in both phases and
(ii) those which occur only in the broken phase.

We consider these in turn:

(i) In the unbroken phase our states are exact chirality
eigenstates. The only perturbative processes which change
chirality are those involving Higgs particles, like that
shown in Figure 3, in which a left-handed  quark
flips chirality when it scatters off a gluon and emits
a Higgs particle. Since we are  taking  only the Higgs-top
Yukawa coupling to be non-zero
this induces a coupling only between left-handed top
quarks, right-handed top quarks and Higgs particles.
The gluon cancels out when we subtract particles from
anti-particles because it is its own anti-particle.
The only other processes changing the number
of particles of a given chirality in the unbroken phase
are the strong anomalous processes, and the weak anomalous
processes which are responsible for the baryon
production. The former couple left-handed quarks
of all flavors to right-handed quarks of all flavors directly; the
latter
couple all the left-handed quarks to all the left-handed leptons.

(ii)
In the broken phase all the processes in the unbroken phase
(except the baryon number violating
processes) are still present, but there are
many additional processes
which couple particles on a given branch of our dispersion
relations to other particle states, because of the
mixing of helicity and chirality states when
the mass is non-zero.

When we calculate interactions between particles in these
states we find that they can couple to one another
through a `helicity-flipping'  gauge boson
exchange. This is also an example of what we called
a `hypercharge violating' process: if we identify the
ingoing states by the hypercharge of the state they
deform into in the unbroken phase, hypercharge is not
conserved.
We have evaluated the  rate for this process in Appendix B.
There are many other such processes e.g. involving
Higgs fields. There are also many flavor changing
processes mediated by $W$ bosons, but these are
zero in the approximation that only the top quark Yukawa
coupling is non-zero.

We now make three further simplifications.
Firstly we will assume that the baryon
number violating  processes can be
neglected, except in their role
as the source of net baryon number. We will
see below that the condition that this be true
is (see also \JPTlonga )
\eqn\condsphal{ \Gamma_{s} < {v_w^2 \over D}}
where $\Gamma_{s}$ is the rate for electroweak
anomalous processes. ${v_w^2/ D}$ is just the
rate of capture of a diffusing particle by
the advancing wall (the inverse of the
time it spends diffusing before capture).

The second further simplification we make
is to neglect completely the  decay processes
in the unbroken phase involving Higgs particles
on the external legs. The  reason we do
this is found in the study of these processes
in \JPTlonga  - in the unbroken phase the equations
\diffdecay\ are precisely the same diffusion-decay
equations obtained there. The change to the results
when these decay processes were incorporated was found
to be a minor numerical one,
and we assume that the same will be true here.
In short the reason is that these processes
do not drive the quantity sourcing baryon number (i.e.
left-handed fermion perturbations) to zero. In
the limit where they are fast enough to
equilibrate locally they simply lead to
a redistribution of particles amongst the
left-handed fermions, right-handed fermions and
Higgs particles.

The final simplifying assumption we make is to take
the diffusion constant of left-handed and right-handed quarks
to be equal. Although this is a very good approximation
since the diffusion properties are dominated by the
strong interactions, it is one which must be treated
with caution in certain limits of extreme suppression
of baryon production by the decay processes. This is
discussed in section 6 of \JPTlonga\ and the treatment given
there can be applied to the present case.
We will not discuss this here.

With these assumptions we can now greatly simplify
the full set of coupled equations \diffdecay\
for all species in the plasma. Defining the variables
\eqn\prev{ \bar \mu_t = N_c (\bar \mu_{t_L} - \bar \mu_{t_R}) \qquad
          \bar \mu_{\Delta}  =
N_c \Sigma_i (\bar \mu_{q_L}^i  - \bar \mu_{q_R}^i)
\qquad \bar \mu_{B}  = N_c \Sigma_i (\bar \mu_{q_L}^i  + \bar
\mu_{q_R}^i) }
where $\bar \mu_{t_L}$ means the difference in the chemical
potential of the left-handed top and
its (right-handed) anti-particle  etc., the sum is over
flavors and $N_c$ is the number of colors, we can extract
a simple set of three coupled equations:
\eqn\coreeqns{\eqalign{
D{\bar \mu_t ^{''}\over T_o}
+  v_w {\bar \mu_t  ^{'} \over T_o}
- \Gamma_f {\bar \mu_t  \over T_o} - {\Gamma_{ss}\over N_f}{\bar
\mu_{\Delta}  \over T_o}&=
- {D \over b} F'(z)  - 2N_c \Gamma_f v_w g_A {Z \over T_o} \cr
D{\bar \mu_{\Delta} ^{''}\over T_o}
+  v_w {\bar \mu_{\Delta}  ^{'} \over T_o}
- \Gamma_f
{\bar \mu_t  \over T_o} -  \Gamma_{ss}{\bar \mu_{\Delta}  \over T_o}&=
- {D \over b} F'(z)  - 2 N_c \Gamma_f v_w g_A {Z \over T_o} \cr
 D{\bar \mu_B ^{''}\over T_o}
+  v_w {\bar \mu_B  ^{'} \over T_o}
- \Gamma_s {\bar \mu_B  \over T_o}
=  +\Gamma_{s}{\bar \mu_{\Delta}  \over T_o} \cr  } }
where
$F = 4 cN_c v_w g_A {( Z m^2)' \over T_0^3}$.
$N_f$ is the number of
fermion flavors - and we see that the first two equations
reduce to a single one in the case $N_f=1$.
$\Gamma_{f}$, $\Gamma_{ss}$ and $\Gamma_{s}$ are the rates
for the helicity-flipping, strong anomalous processes
and weak anomalous processes respectively.
With this new convention equations
\coreeqns\ take the form of diffusion equations corresponding
to those
in the companion paper \JPTlonga, and the rates $\Gamma$ are identical
to
the rates used there.
$\Gamma_{ss}$ and $\Gamma_{s}$ can be read off
from  \JPTlonga, and the helicity
flip rate is derived in Appendix B:
\eqn\ratesfinal{\eqalign{ \Gamma_f \approx &
{1\over 2} {m_t^2\over T^2}\alpha_s^2 T
 \approx {1\over 100}{m_t^2\over T}\cr
\Gamma_{ss} = & {32 N_f \over 3} \kappa_{ss} \alpha_s^4 T
\approx \kappa_{ss} {T \over 40 } \cr
\Gamma_{s} = & 9N_F \kappa_{s} \alpha_w^4 T \approx \kappa_{s} {T \over
3\times 10^4}\cr
}}
where ${8 \over 3} \kappa_{ss} \alpha_s^4 T^4 $
and $ \kappa_{s} \alpha_s^4 T^4 $ are the number
of strong and weak (respectively) anomalous processes
per unit volume per unit time
$N_F$ is the number of fermion families.

Before proceeding to analyze these equations and calculate
the baryon asymmetry, we stop and review the numerous
assumptions we have made in deriving the equations
\coreeqns.

$\bullet$ Assumption 1: $L > {1 \over T} $ so that most
particles in the plasma are indeed accurately described
by the WKB approximation. Typical wall thicknesses
are $L \sim 20/T$.

$\bullet$ Assumption 2:$v_w < { L \over 3D}$, the `thermalization'
condition for the applicability of our fluid ansatz. We calculated
$D \approx 5/T$, so for typical thick wall $L \sim 20/T$ this
is an extremely good for any wall velocity.

$\bullet$ Assumption 3: $v_w < {1 \over \sqrt{ 3} }$. We work to linear
order in the wall velocity assumed smaller than the speed of sound
in the plasma, so that perturbations can propagate into the region
in front of the wall and give `non-local' baryogenesis. This
restricts us to modest wall velocities.

$\bullet$ Assumption 4: $\Gamma_s < {v_w^2/ D}$, so that
the back-reaction of the baryon number
violating processes on the perturbed quantities
can be neglected.
Using \rates\
we see that this corresponds to $v_w > 10^{-2} \sqrt{\kappa_s}$.
Numerical simulations indicate $\kappa_s \sim 0.1 - 1$
so this is very consistent with favored wall
velocities $v_w \sim 0.1 - 1$.
If $\Gamma_s > {v_w^2/ D}$ the sphaleron rate can equilibrate
in front of the wall.

$\bullet$ Assumption 5: All decay processes involving
Higgs particles on the external legs can be neglected.
This can be revised along
the lines discussed in \JPTlonga, and should lead only to minor
numerical corrections.

$\bullet$ Assumption 6: The diffusion constants of quarks
of opposite chirality can be taken to be  equal.
A very good approximation which
need only be revised for the limit of extreme suppression
of an asymmetry by decay processes.

$\bullet$ Assumption 7: Only the top-quark Yukawa
coupling is  non-zero, so we do not describe the case
of a two doublet model in which the lepton Yukawa coupling
may be large.

{\bf 6. The Baryon Asymmetry}

The baryon number density
$B= {1\over N_c}{T^2 \over 12} \bar \mu_B$,
and  calculating  with the last equation in \coreeqns\
we find its value at the front of the wall to be
\eqn\baryonsb{  B_o = {1 \over N_c}{T^2 \over 12} \bar \mu_B
  \approx  {1\over N_c}
{ T^2 \over 12}\times {\Gamma_s  \over v_w} \int _{L \over 2} ^\infty
\bar \mu_{\Delta} (z) dz.}
This is the final value of the baryon number in the
broken phase under the assumption that the baryon number
violating processes are turned off everywhere  in the
broken phase. The expression \baryonsb\ is valid with
the assumption we made that $\Gamma_s$ satisfies
${ v_w^2 \over  \Gamma_s D  }>> 1$.

To determine the final baryon asymmetry we must
use the first two equations in \coreeqns\ to extract
$\bar \mu_\Delta$ in front of the wall
which is the effective source for the baryon production.

We will not solve exhaustively the two coupled equations
for $\bar \mu_t$ and $\bar \mu_\Delta$ in \coreeqns, but limit
ourselves to the case that the dominant decay process
is $\Gamma_{ss}$ so that we can drop the term
$\Gamma_f \bar \mu_t$ in the second equation in \coreeqns.
This is justified for all $\kappa_{ss}$ in the range  $0.1 - 1$, since
$m_t^2< T^2/4$ on the wall ({\it cf.\/} {\ratesfinal\/}).
We are then  left with a single equation for $\bar\mu_{\Delta}$:
\eqn\corered{\eqalign{
D{\bar \mu_\Delta  ^{''}\over T_o}
+  v_w {\bar \mu_\Delta  ^{'} \over T_o}
- \bar \Gamma_\mu  {\bar \mu_\Delta  \over T_o} =
- {D \over b} F'(z)  - 2N_c \Gamma_\mu  v_w g_A {Z \over T_o}  } }

To solve \corered\ we need to find particular solutions
as well as solutions to the homogeneous equation.
The latter are $\sim e^{-\lambda z}$
where
\eqn\rootsb{ \lambda=
\cases { \lambda_D= {v_w \over D}(1 +{D \bar \Gamma_\mu \over  v_w^2 }
+..)
\quad {\rm and} \quad -\lambda_d =-{\bar \Gamma_\mu \over v_w}
(1- {D\bar\Gamma_\mu \over  v_w^2 }+..)
&${D\bar\Gamma_\mu \over v_w^2} <<1$   \cr \pm \lambda_\pm = \pm
\big [( \sqrt{{\bar \Gamma_\mu \over D} }+...) \mp {v_w \over 2D}\big ]
&          ${D\bar\Gamma_\mu \over v_w^2} >>1$ \cr } }
The behaviour of the solutions is determined by
the parameter  ${D\bar\Gamma_\mu \over v_w^2}$
(precisely as in the companion paper  \JPTlonga).
It characterizes the competition between decay and diffusion,
${D \over v_w^2}$ being the time a typical particle spends diffusing
in front of the wall before being caught.
When   ${D\bar\Gamma_\mu \over v_w^2} <<1$, decay becomes
irrelevant in front of the wall,
and the only effect of the decay
processes is to restore thermal equilibrium far behind the wall.
This is also the criterion we need to use to see if
any process is of relevance to the problem we are considering
- in particular we used it above in \condsphal, and implicitly
in assuming that the Yukawa couplings which we have set to
zero mediate decay processes which are slow in precisely this
sense.

We now consider the solution of \corered\ for each of the
two source terms on the right hand side separately.

{\bf 6.1 Classical Force Baryogenesis }

We return again to the `ramp' ansatz \ansatza, with $A=4cN_c$ ,
and take only the force term as a source in \corered.
The problem is now homogeneous with boundary conditions
\eqn\bcsforce{\eqalign{
D ({\bar \mu_\Delta \over T_o})^{'}+ v_w  ({\bar \mu_\Delta \over T_o})
 \big | ^{\pm{L \over 2} + \epsilon }_{\pm {L \over 2} - \epsilon }=
\pm {DF_o\over b} \qquad
D ({\bar \mu_\Delta \over T_o})
 \big | ^{{L \over 2} + \epsilon }_{{L \over 2} - \epsilon }= 0  \qquad
D ({\bar \mu_\Delta \over T_o}) \big | ^{-{L \over 2} + \epsilon }_{-{L
\over 2} - \epsilon }= {DL F_o \over b}   } }
derived by integrating \corered\ with the assumption
that the integral of $\bar \mu_\Delta$ is continuous across the
boundaries. Requiring $\bar \mu_\Delta$ to be finite
at $\pm \infty$ we take
\eqn\form{ \bar\mu_\Delta = \cases { A_f  e^{-\lambda_f z} & $z>
{L\over 2}$ \cr
 A_w e^{-\lambda_f z} +  B_w e^{\lambda_b z} & $-{L\over 2}< z< {L\over
2}$ \cr
 B_b e^{\lambda_b z} & $z< -{L\over 2}$  \cr }}
where $A_f$, $A_w$, $B_w$ and $B_b$ are constants, and from \rootsb\
we have
\eqn\rootsc{ \eqalign{
\lambda_f &= \cases { \lambda_D & ${D\bar\Gamma_\mu \over v_w^2} <<1$
\cr                      \lambda_+ & ${D\bar\Gamma_\mu \over v_w^2} >>1$
\cr } \cr \lambda_b &= \cases{\lambda_d  & ${D\bar\Gamma_\mu \over
v_w^2} <<1$ \cr
                      \lambda_-  & ${D\bar\Gamma_\mu \over v_w^2} >>1$
\cr} \cr
 }}
Using the boundary conditions we determine
the solution in front of the wall to be
\eqn\forcesolna {-{1\over b}F_o \big [ {1 \over (\lambda_f+\lambda_b)L}
-{ (\lambda_b+\lambda_D^o)L \over (\lambda_b+\lambda_f)L}
 e^{-\lambda_f L} \big ]
e^{-\lambda_f(z-{L \over 2})} }
where $\lambda_D^o= {v_w \over D}$.
In various limits this reduces to
\eqn\forcesolnb{v_w{A\over b}{ Z_o m_o^2 \over 2 T_o^3} \cases {
  {v_w L \over D}e^{-{v_w \over D}(z-{L\over 2})} &
${ v_w L \over D}<<1 \quad{\rm and}\quad{D\bar\Gamma_\mu \over v_w^2}
<<1\quad {\rm }$ \cr
  {2 D\over v_w L} e^{-{v_w \over D}(z-{L\over 2})}&
${ v_w L \over D}>> 1 \quad{\rm and}\quad{D\bar\Gamma_\mu \over v_w^2}
<<1
\quad {\rm }$ \cr
  {1\over 2}\sqrt {\bar \Gamma_\mu \over D} L e^{-\sqrt {\bar
\Gamma_\mu \over D}(z-{L\over 2})} &
$\sqrt {\bar \Gamma_\mu \over D}L <<1 \quad{\rm
and}\quad{D\bar\Gamma_\mu \over v_w^2} >>1
\quad {\rm }$ \cr
  { 1 \over L} {\sqrt {D \over \bar \Gamma_\mu }} e^{-\sqrt {\bar
\Gamma_\mu \over D}(z-{L\over 2})}&
$\sqrt {\bar \Gamma_\mu \over D}L >> 1 \quad{\rm
and}\quad{D\bar\Gamma_\mu \over v_w^2} >>1
\quad {\rm }$ \cr }}
where $Z_o = Z(-{L\over 2})$ and $m_o = m(-{L\over 2})$.

The first two cases agree precisely with what we saw when we analysed
the force neglecting decay processes in section 5.2.
The prefactor in \forcesolnb\
corresponds to the average of the solution approached
on the wall for ${v_w L\over D}<<1$, but with the opposite sign
so that the integrated contribution in front of the wall
cancels that on the wall.
The two limits for ${v_w L \over D}$ are the limits
of good and poor transport which we discussed.

The two other cases given in \forcesolnb\
tell us how these solutions are modified when
${D\bar\Gamma_\mu \over v_w^2} > 1$.
The penetration of the diffusion solutions
into the unbroken phase is reduced since
$\sqrt {\bar \Gamma_\mu \over D} > {v_w \over D}$.
As we noted this is just the condition that the
average diffusing particle's time in the unbroken
phase before capture be longer than its
decay lifetime.
A second parameter enters in determining how the
amplitude of the solution in front of the
wall is changed. In the first case,
$\sqrt { {\bar \Gamma_\mu \over  D} }L << 1$,
the amplitude in fact compensates by increasing
so that the integrated
result $\int_{{L\over 2}}^\infty \bar\mu_\Delta$
is unchanged (up to a factor of two).
In the second case, $\sqrt { {\bar \Gamma_\mu \over  D} }L >> 1$,
the amplitude is attenuated and the integrated result
differs from that in the no-decay case by
a factor of  $v_w^2/2 D\bar \Gamma_\mu<< 1$.
The physical meaning of the
parameter $\sqrt { {\bar \Gamma_\mu \over  D} }L$
is also simple. It is (the square root of) the ratio
of the time a particle
takes to diffuse across the wall to its decay lifetime.
So what the the third and fourth cases in \forcesolnb\
tell us is that the net density of particles in front
of the wall in these stationary solution is not
changed (up to a factor of two) unless the decay
process is fast enough so that particles can decay
as they cross the wall. This is a surprising result
as one might expect the only relevant parameter to be
${D\bar\Gamma_\mu \over v_w^2}$, which compares the
decay time to the time a particle actually spend in
front of the wall.

The factor of two has a simple explanation which will be familiar to
the close reader of \JPTlonga. If one considers a diffusion/decay
equation for a single species with a given injected flux
modeled by a delta function, one finds  two
distinct regimes corresponding to the value
of ${\Gamma D \over v_w^2}$. In the high velocity diffusion
regime (${\Gamma D \over v_w^2}<<1$) the stationary solution
puts most of the injected flux
into the amplitude
in front of the wall; in the low velocity case
the solution has equal and opposite amplitude in front and
behind, sharing the injected flux so that the
amplitude is exactly half that in the first case.

One might worry that some of these features are
artefact of our ansatz \ansatza\ in which there is
a discontinuity in the profile of $m^2 Z$  at the back
of the wall. To check this we take
 instead the `bell' ansatz:
\eqn\bell{ Av_w {Z m^2\over T_o^3}= \cases { {F_o L\over 4}
    ( 1 -{ 4 z^2 \over L^2} )  &
          $-{L\over 2} < z < {L\over 2}$  \cr
 0 & \rm {otherwise} \cr }     }
and using the same conventions as in \form\ we find
in front of the wall
\eqn\forcebella{- {1\over b}F_o  {1 \over (\lambda_f+\lambda_b)}\big [
{2\lambda_b \over \lambda_d^o} {1 \over \lambda_D^o L}
(1- e^{-\lambda_fL}) - (1+ e^{-\lambda_f L}) \big ]
e^{-\lambda_f(z-{L \over 2})} }
where $\lambda_d^o= {\bar \Gamma_\mu \over v_w}$
and $\lambda_D^o= {v_w \over D}$. In
the same limits as before this reduces to
\eqn\forcebellb{{ F_o \over 6b } \cases {
  { v_w L \over D}e^{-{v_w \over D}(z-{L\over 2})} &
${ v_w L \over D}<<1 \quad{\rm and}\quad{D\bar\Gamma_\mu \over v_w^2}
<<1
\quad {\rm }$ \cr
  {6D \over v_w L} e^{-{v_w \over D}(z-{L\over 2})}&
${ v_w L \over D}>> 1 \quad{\rm and}\quad{D\bar\Gamma_\mu \over v_w^2}
<<1
\quad {\rm }$ \cr
  {1 \over 2  }\sqrt {\bar \Gamma_\mu \over D} L e^{-\sqrt {\bar
\Gamma_\mu \over D}(z-{L\over 2})} &
$\sqrt {\bar \Gamma_\mu \over D} L <<1 \quad{\rm
and}\quad{D\bar\Gamma_\mu \over v_w^2} >>1
\quad {\rm }$ \cr
  {3 \over L}\sqrt {D\over \bar \Gamma_\mu } e^{-\sqrt {\bar \Gamma_\mu
\over D}(z-{L\over 2})}&
$\sqrt {\bar \Gamma_\mu \over D} L >> 1 \quad{\rm
and}\quad{D\bar\Gamma_\mu \over v_w^2} <<1
\quad {\rm }$ \cr }}
This again shows the same results in each case. The only
difference is the numerical factor which comes from
averaging the profile over the wall, which is precisely what
we anticipated since ${F_o \over 6} = <Av_w {Z m^2 \over T_o^3}>$.

Using \baryonsb\ we now finally calculate the baryon asymmetry
in its standard  form and find
\eqn\bauone{ {n_B \over s}=  {4 \over  g_*}  \kappa_s
 \alpha_w^4  \int {g_A Z m^2 \over T_o^2} dz
 \cases { 1 & if $v_w^2 > \Gamma_{ss} D \approx  {\kappa_{ss}\over 6}$
\cr
{1\over 2} & if $ v_w^2 <  { \kappa_{ss} \over 7}$  and
$\sqrt { { \Gamma_{ss} \over D} } L \approx  { \sqrt {\kappa_{ss} } L T
\over 14} < 1$ \cr
{\eta D \over \Gamma_{ss} L^2 } &   if $ v_w^2 <  {\kappa_{ss}\over 7}$ and
${ \sqrt {\kappa_{ss} } L T \over 14} > 1$ \cr } }
where $s={2 \pi^2 \over 45} g_* T^3$ is the entropy density of
the universe, $g_*$ the number of relativistic degrees of freedom,
and $L$ the thickness of the wall.  $\eta$ is a geometrical
factor which must be calculated for the particular wall profile
($1$ for `ramp', $3$ for `bell').
In all cases we have assumed $\Gamma_s<<v_w^2/D$; in the case of a
wall moving sufficiently slowly that this condition is violated, one would
recover $n_B\propto v_w$ as expected.
Recall that these results are derived
under the assumption that the strong sphaleron is
the dominant decay process everywhere. Other cases can be treated
using the more general form of \coreeqns.

This  expression is remarkably  simple.
Most strikingly $L$, $v_w$ and $D$ all cancel out in
the answer in the most interesting (and quite plausible)
regimes.
For typical `slow thick wall' parameter values e.g.
$L \sim {10 \over T}$, $v_w \sim 0.1$,
$\kappa_{ss} \sim 0.1-1$ and $m_t \sim T$, the conditions
for our derivation hold.
The result is
 $\approx 2 \times 10^{-2} \kappa_s \alpha_w^4 ({m \over T})^2 \Delta
\theta
\approx 2 \times  10 ^ {-8} \kappa_s \Delta \theta$,
where $\Delta \theta \equiv m_t^{-2} \int m_t(z)^2 g_A Z dz $
is a measure of the $CP$ violating condensate
on the wall.

The magnitude of $\Delta \theta$
depends on  the precise profile
of the wall, with the greatest effect occurring
(in the two doublet theory) if
the phase $\theta$ rolls fastest where the mass is large.
In two doublet theories it will have the same sign on every bubble
in a way determined by the effective potential, and
can be $O(1)$  consistently with measurements of CP violation.
In the case of the standard model $Z$ condensate
this factor will contain a suppression (potentially many
orders of magnitude) depending on exactly how one sign
of the condensate comes to dominate over the other. A full
treatment of this case is required which goes beyond
the scope of this paper
\ref\ntprep{S. Nasser and N. Turok, in preparation}.

{\bf 6.2 Local Spontaneous Baryogenesis }

In the previous section we have concentrated primarily on the
chemical potential fluctuations in front of the  wall where the
sphaleron rate
is unsuppressed. This  clearly dominates baryon production in the case of
efficient transport ${v_w L \over D} < 1$.
In the case of inefficient transport (when the wall is very thick and slow)
so  that ${v_w L \over D} >> 1$ the non-local baryogenesis will be suppressed
by $\sim {D \over  v_w L}$ and local baryogenesis may dominate.
We analyse this case now to relate our treatment to that in the literature
(in particular in CKN's work \cknsba)
prior to \JPTtran, in which the potential importance of transport
in the plasma was noted.

When we combine the first two equations in \coreeqns\
and ignore transport ($D=0$), we obtain
\eqn\notransport{v_w \mu ''_\Delta - (\Gamma_f +\Gamma_{ss}) \mu'_\Delta +
{\Gamma_f \Gamma_{ss}\over v_w} (1-{1 \over N_f}) \mu_\Delta
=-2 N_c \Gamma_f v_w g_A Z'}
The force term source drops out in this limit simply
because in order to induce
perturbations particles must move in response to the force,
(and $D=0$  ``freezes'' the particles).
The case of the perturbations induced by the second source terms
on the right of \coreeqns\ is quite different. The effect of the
field $Z$ is local - it creates a perturbation at the
point at which it is turned on by changing the interaction
rates of hypercharge violating processes.

For comparison with the previous literature it is instructive to consider
the following cases in which the solutions to this equation can be
read off simply.

$\bullet$ {\it Case 1:\/} $\Gamma_{f}^{-1}<< L/v_w << \Gamma_{ss}^{-1}$
for which
\eqn\notransportone{
\mu_\Delta = 2 N_c v_w g_A Z
}
to leading order in ${\Gamma_{ss} \over \Gamma_f}$.

$\bullet$ {\it Case 2:\/} $\Gamma_{ss}^{-1},\;  \Gamma_{f}^{-1} << L/v_w $
for which
\eqn\notransporttwo{
\mu_\Delta = -2 {N_c \over 1-N_f^{-1}} {v_w\over \Gamma_{ss}} v_w g_A Z'
}
Here we ignore the homogeneous solutions which simply describe
how the perturbations induced on the wall decay away behind the
wall.

The first case gives us what would be obtained by finding the
local thermal equilibrium subject to the constraints
imposed by the interactions locally, neglecting strong sphalerons.
This is precisely
the limit calculated by CKN in \cknsba, albeit with
a fermionic hypercharge in place of $v_w Z$ and
correspondingly a fermionic hypercharge violating
process in place of $\Gamma_f$. The requirement
$\Gamma_{f}^{-1}<< L/v_w $ is
just the condition that the interaction time for
the hypercharge violating process be short in
comparison to the time of passage of the wall,
a requirement imposed by CKN on
the fastest fermionic hypercharge violating process.

 From the second case we see that in if one
takes $\Gamma_{ss} \rightarrow \infty$ i.e. puts
the strong sphalerons  into equilibrium,
the result is zero. This is a simple way of seeing
the result obtained  by Giudice and Shaposhnikov  in
\ref\gs{G.F. Giudice and M. Shaposhnikov,
Phys. Lett. {\bf B326}, 118 (1994).}.
The result \notransporttwo\ tells us how to calculate corrections
to this constrained local equilibrium calculation, keeping
the constraints but taking the rates to be finite,
and gives what one might guess: the result
is approximately equal to the equilibrium result of
\notransportone\ with an additional suppression
${v_w \over \Gamma_{ss} L}$.

These formulae also show - as remarked earlier
and originally pointed out by Dine and Thomas
in \dt\ -
that the perturbation is not well modeled by
a fermionic hypercharge potential. But is it a potential
for total hypercharge? In the rest frame
of the plasma, with our assumption of a
stationary wall profile, the time
component of $\tilde{Z}_\mu$ is $v_w Z$.
But one must be careful about the conclusion
that one simply replaces $v_w Z$ by
whatever this time  component is. One evident problem is
that one loses the explicit $v_w$ dependence, as such a potential
can in principle be non-zero with the wall at rest,
and the result when $v_w \rightarrow 0$ is then non-zero.
The problem is exemplified in CKN's new calculation
of spontaneous baryogenesis \cknsbnew\ which incorporates
the effect of transport through a diffusion equation,
in which only the time component of $\tilde{Z}_\mu$ appears.
However if we consider the case in which {\it only} the
time component of $\tilde{Z}_\mu$ is nonzero,
their diffusion equation does not approach the correct thermal
equilibrium as $v_w \rightarrow 0$. To recover this
must include the force term which
results in a net overdensity in proportion
to $\tilde{Z}_o$. In general, a correct treatment should include
both space and time components of the gauge field, and the
corresponding force terms. In our case, where
the field configuration is assumed static in the wall frame,
one can work in this frame and in this case the only
non-zero component is the {\it spatial} $z$-component, and the
time component plays no role.
We will not compute in this case the baryon asymmetry,
as it involves making specific assumptions about how
the electroweak sphaleron rate behaves on the wall.
We concentrate instead on the non-local variant of
this mechanism.

This brings us to one final remark. When the
transition proceeds, as we have assumed, by bubble
nucleation, $\tilde{Z}_\mu$ is a spacelike vector if we take
the wall profile to be stationary.
The one case in which it can be modeled consistently
as a timelike total hypercharge potential is when the
transition occurs by spinodal decomposition, where the
Higgs fields all roll together in the same way everywhere
in space. Going back to our dispersion relation
in section 2 for this case we would follow through our
derivations in the same way. We would need to redefine
the chemical potential by $g_A \tilde{Z}_o$ and, because of the spatial
homogeneity, would find no force term, and for
the same reason, would discard the spatial gradients
and, with certain assumption about how the transition
proceeds, the time derivatives too. We would arrive
at \notransportone\ above with $v_w Z$ replaced by $\tilde{Z}_o$.

\vfill\eject

{\bf 6.3 Non-Local Spontaneous Baryogenesis }

We now turn to the case where this local effect
is turned into a non-local one by the effects of transport.
Perturbations can then be generated in front of the wall
where the sphaleron rate is unsuppressed, in contrast to
the local effect which has the problem that
baryons must in the region where the effect which
is  vev squared suppressed is turned on.
For a discussion of this effect we also point the reader
to \cknsbnew.

We turn again to \corered, again assuming $\Gamma_{ss}$
to be the dominant decay process, taking the second source term
and the `ramp'  ansatz:
\eqn\ansatzb{ {\Gamma_f Z \over T_o} = \cases {  -{\Gamma_o Z_o\over
T_o} ({z-{L\over 2} \over L})  &
          $-{L\over 2} < z < {L\over 2}$  \cr
 0 & \rm {otherwise} \cr }     }
The vev squared dependence in the rate is absorbed in
the ansatz.
With this ansatz $\bar \mu_\Delta$ is then continuous
everywhere but the equation is not homogeneous.

The particular solution for this ansatz is
\eqn\nlsponps{ {\bar \mu_\Delta \over T_o}= \cases {
-v_w{\Gamma_\mu \over \bar \Gamma_\mu} {g_A Z_o \over T_o}
\big [ { z-{L\over 2}\over L} +{v_w \over \bar \Gamma_\mu L} \big ]  &
                 $-{L\over 2} < z < {L \over 2}$ \cr
0 & ${\rm otherwise}$ \cr}}
and the solution in front of the wall
\eqn\nlsponsoln{ -2N_cv_w{\Gamma_\mu \over \bar \Gamma_\mu} {g_A Z_o
\over T_o}
{1 \over (\lambda_f+\lambda_b)L} \big [
(-1 + { \lambda_b \over \lambda_d^o})
+(1-{\lambda_b \over \lambda_d^o} +\lambda_b L)e^{-\lambda_fL} \big ]
e^{-\lambda_f(z-{L \over 2})} }
using the same
conventions again as in \form, and $\lambda_d^o= {\bar \Gamma_\mu \over
v_w}$.

Again we simplify this in various limits to
\eqn\nlsponsolnb{ +2N_cv_w{ \Gamma_\mu \over \bar \Gamma_\mu} {g_A Z_o
\over T_o}
\cases { -{1\over 2}{ v_w L \over D}({D \bar \Gamma_\mu \over  v_w^2})
e^{-{v_w \over D} (z-{L\over 2})} &
$ { v_w L \over D}<< 1 \quad{\rm and}\quad{D\bar\Gamma_\mu \over v_w^2}
<<1\quad {\rm  }$ \cr
 {D \over v_w L }({D \bar \Gamma_\mu \over v_w^2})
e^{-{v_w \over D}(z-{L\over 2})} &
${ v_w L \over D}>> 1 \quad{\rm and}\quad{D\bar\Gamma_\mu \over v_w^2}
<<1
\quad {\rm  }$ \cr
  {1 \over 4  }\sqrt {\bar \Gamma_\mu \over D} Le^{-\sqrt {\bar
\Gamma_\mu \over D}(z-{L\over 2})} &
$\sqrt {\bar \Gamma_\mu \over D} L <<1 \quad{\rm
and}\quad{D\bar\Gamma_\mu \over v_w^2} >>1
\quad {\rm  }$ \cr
  {1 \over 2  L}\sqrt {D \over \bar \Gamma_\mu} e^{-\sqrt {\bar
\Gamma_\mu \over D} (z-{L\over 2})}&
$\sqrt {\bar \Gamma_\mu \over D} L >> 1 \quad{\rm
and}\quad{D\bar\Gamma_\mu \over v_w^2} >>1
\quad {\rm  }$ \cr
   }}
These solutions are very similar to those in the case of
the force, the amplitude being of the form of
the amplitude of a `static' solution \notransportone\
on the wall
multiplied by a factor which depends on the relative
importance of decay and diffusion. The `static'
solution in this case is attained in the limit that
there is no transport and the decay processes are turned on.

In contrast, the solution
\vlimit\ for the force is reached when transport is perfect
and the decay processes are turned off.

For the `spontaneous' effect,
we see that as $D\rightarrow 0$
the amplitude  in front of the wall
vanishes rapidly ($\sim D^2$).

As $D\rightarrow \infty$ the amplitude goes as ${1\over \sqrt{D} }$
but the tail integrates to give exactly half the
solution \notransportone, but in this case not with the opposite
sign. The arguments which we used to explain
the sign in the case of the force do not apply as they rested on
\total. In the `spontaneous' result
the sign is in all but the first case the same as
that of the `static' solution. What is happening is that the
overdensity which the local process creates diffuses out to
the region in front of the wall. Only in the first
case, where the transport
is much more efficient than the decay  do particles diffuse in
to `cancel' the density on the wall.

The baryon to entropy ratio can again be calculated.
We find precisely the same expressions in
\bauone\ but with the replacement
\eqn\replace{  \int {g_A Z m^2 \over T_o^2} dz
\rightarrow {b \over 2c} \int {\Gamma_f g_A Z \over  \Gamma_{ss}} dz
  \cases { {D \bar  \Gamma_{ss} \over  v_w^2}\approx {D   \kappa_{ss}
\over 7 v_w^2}&
 $ v_w^2 >> {\kappa_{ss} \over 7}$ \cr
1 &   $ v_w^2 << {\kappa_{ss} \over 7}$ \cr }}
Using $\Gamma_f$ and $\Gamma_{ss}$ from  \ratesfinal\
so that $\Gamma_f/\Gamma_{ss}\simeq 0.4 (m_t/T)^2/\kappa_{ss}$,
this result is seen to be equal to the force result
multiplied by
$\simeq 1/50\kappa_{ss}\alpha_s^2\simeq 1/\kappa_{ss}$.
(This is valid for $v_w^2 << {\kappa_{ss} \over 7}$
and, as we assumed, $\Gamma_{ss} >> \Gamma_f$.)
The final result is that the two effects have quite
different parametric dependence but appear to be
roughly of the same order of magnitude.

A striking difference between the classical force effect and the
spontaneous baryogenesis effect is
in the dependence on the strong sphaleron rate.
In the latter there is an inverse dependence on
the sphaleron rate for a sufficiently large $\kappa_{ss}$.
This dependence has been noted by CKN in their numerical
study \cknsbnew. In contrast,
the force sourced result is not
suppressed by the strong sphalerons if
${\sqrt{\kappa_{ss}}L T/14}<1$.

\medskip

\centerline {\bf 7. Comparison of Thin and Thick Wall Regimes  }

It is interesting to compare the results in this paper with
those in \JPTlonga, where we considered the case of
baryogenesis produced
by reflection off a thin wall.
The two calculations  are most easily compared
by looking at the version of the fluid equations \coreeqns.
Once we established that the thermal
fluctuations were unimportant we were able to reduce the system
to a  diffusion equation precisely analogous to that in
\JPTlonga, but including a  force term
which extends over a finite region of space - the wall -
where we had before the derivative of a delta function,
modeling the injected flux.
\eqn\replaceb{ \xi J_o \delta \rightarrow {D \over b} F(z)}

In the classical force case,
integrating the right hand side gives zero because the
force is derived from a potential. Unlike the
injected case, the
effect of the classical force vanishes as the wall thickness
is taken to  zero.
This is because the force simply
speeds up and slows down particles that pass over it.
We have not included the effect of the force as a reflecting
barrier on low momentum states, since we neglected the
effect of the `flow' of one  branch of the dispersion relation
into the other at low momentum. If we include this
we should obtain in the thin wall limit what one would
calculate for the reflected flux in the WKB limit.
The quantum mechanical reflection we can of course not
recover.
To compare the magnitude of the classical force effect
with the WKB reflection effect
we take the ratio of
the amplitudes of the diffusion tails at the front
of the wall. Doing this we find,
neglecting decay and assuming efficient transport,
\eqn\compare{ { \mu_{refl} \over \mu_{force}} \sim
{\xi \over D} L T{D \over v_w L }  \eta  }
The most striking result is the the very different parametric
dependence.

The effect of introducing a decay term - in particular
due to strong sphalerons - is quite different in the two cases.
In our  thin wall calculations \JPTlonga\
the suppression which resulted
was ${v_w \over \sqrt{\Gamma_{ss} D}}$ for ${v_w^2 \over
\Gamma_{ss}D}<<1$,
and this same suppression is seen in the nonlocal spontaneous effect.
In the case of the classical force this
 suppression (which comes from the shortening of the diffusion tail
in front of the wall)
 is compensated for by an
 increase in amplitude until $\Gamma_{ss}$ enters the regime
 $\sqrt{\Gamma_{ss}  / D}L > 1 $.
An explanation may be found  from the equation for $\bar \mu_\Delta$,
 \coreeqns.
Dropping both $\Gamma_f$ terms, integrating once one finds that
the integral of $\bar \mu_\Delta$ over all space must be  zero.
What the condition $\sqrt{\Gamma_{ss}  \over D}L <1$ means is
that strong sphalerons have little effect on the wall.
But because $\int \mu_\Delta=0$, the compensating tail
in front of the wall is also unaffected. No analogous conservation law
holds in the spontaneous baryogenesis effect, because the whole
effect is driven by a decay process ($\Gamma_f$) and
strong sphaleron suppression occurs.

\medskip

\centerline {\bf 8.  Conclusions }

In this paper we have developed a new procedure
to describe perturbations produced by a $CP$ violating bubble
wall moving through the plasma.

We have shown how a Boltzmann equation can be used to describe the
dynamics of thermal particles in the plasma
when scattering processes are important.
We now conclude with several remarks:

$\bullet$ The treatment relies on a WKB approximation which is good
for most of the particles in the plasma. It does not describe
the effect of the background on low momentum particles for which
this WKB approximation is not valid. A  full treatment
of all the perturbations produced by the propagating wall
remains an open problem - for
recent attempts see \ref\hn{ P. Huet and A. E. Nelson, Phys.Lett. {\bf 355B},
229 (1995) and UW-PT-95-07, hep-ph/9506477; A. Riotto, SISSA-117-95-AP,
hep-ph/9510271.}.

$\bullet$ We have neglected the effect of screening.
The effects of the long range fields may be incorporated through the
appropriate terms in the Boltzmann and fluid
equations.
We have not done this as
it greatly complicates our analysis by coupling all species.
However,
in section 7 of \JPTlonga\ we discuss this issue in some detail.
As illustrated by a simple model calculation
of screening by  quarks and
leptons presented there, we do not expect that the effects
of screening would alter our final results by  a factor very different
from unity.

$\bullet$  There are many improvements and refinements of our
calculation which
are possible. The Boltzmann equation is in principle soluble
without any truncation, and certainly there are other approximations
which can be used.
$\bullet$  One caveat must be added to our justification of
the neglect of the Higgs particles in the determination
of the perturbations driving baryogenesis.
By dropping the Higgs particles we are assuming that
they themselves are not significantly perturbed by a force
term. In the thin wall
case we also assumed that there was no injected flux
in Higgs particles. This is  not justified as the
dynamics of the Higgs particles are likely to be
non-trivial in the background of a
changing vev. It is quite conceivable
that such an effect could be important.

$\bullet$  The methods we have developed should be useful in
approaching the problem of the determination of the
wall velocity and backreaction on the wall due
to reflection. In particular we derived a set of
force terms due to the mass which we did not make use
of in the calculation reported here.

$\bullet$ An important outstanding problem is to relate the techniques
developed here to some formal field theoretic methods such as those used in
\ref\McLerranSTV{L. McLerran, M. Shaposhnikov, N. Turok and M. Voloshin,
Phys. Lett. {\bf B 256}, 451 (1991).}
and more recently in \ref\cpr {D. Comelli, M. Pietroni and  A. Riotto;
preprints hep-ph/95056278, hep-ph/9504265 and references therein. }.

\medskip

\centerline{\bf  Acknowledgements}

M.J is supported by a Charlotte Elizabeth Procter Fellowship,
and the work of T.P.  and N.T. is partially   supported by
NSF contract PHY90-21984, and the David and Lucile Packard
Foundation. We thank the Isaac Newton Institute, Cambridge
for hospitality during the completion of the manuscript.

\bigskip

\centerline {\bf Appendix A. Collision Integrals for
Boson Exchange\footnote{$^\dagger$} {\rm We thank Guy Moore for pointing out
an error in an earlier version of this appendix.}}

We consider first the contribution to the collision integrals in
the fluid equations which come from the $t$-channel gauge boson
exchange processes shown in Figure 2.\footnote{$^\diamond$}{ In this paper
we do not calculate all of $t$ channel tree level diagrams that
contribute to the diffusion constant. We refer the reader to
\wall\ where this has been done systematically. }
We will treat not just the quarks but also the
right and left-handed leptons, primarily
because it is interesting to compare our result for the diffusion
constant with those obtained with our previous method in \JPTlonga.

When we integrate the Boltzmann equation as described in the text
over $\int {d\!\!\!^-}^3 p$, $\int p_o {d\!\!\!^-}^3p$ and
$\int p_z {d\!\!\!^-}^3p$ there are six integrals we need to evaluate:
\eqn\collintexpa{\eqalign{ \int_{p,k,p',k'}
f_p (1-f_p) f_k (1-f_k) |{\cal M}|^2
    (2\pi)^4{\delta}^4 (p+k-p'-k')
(p_o-p'_o) &\times  \cases { 1 \cr p_o \cr p_z } \cr
\int_{p,k,p',k'}
f_p (1-f_p) f_k (1-f_k) |{\cal M}|^2
    (2\pi)^4{\delta}^4 (p+k-p'-k')
(p_z-p'_z) &\times  \cases { 1 \cr p_o \cr p_z }  }}
where we have ignored the higher order terms in the perturbations
$\delta T$, $\bar v$ and $\mu$ so that fermion population densities (in the
plasma frame) are $f_{p_i}=1/(1+\exp (p_i^0/T_0))$, $p_i=\{p,k,p',k'\}$.

For the gluon exchange process the scattering amplitude is
\eqn\mgluon{ |{\cal M}_q|^2= A_q {s^2+u^2\over (t-  m_g^2)^2}  }
where $A_q=32 g_s^4$, $t=(p-p'\,)^2\simeq -2p\cdot p'$,
$s=(p+k\,)^2\simeq -2p\cdot k$,
$u=(p-k'\,)^2\simeq -2p\cdot k'$,
$g_s$ is the strong coupling constant and
$m_g$ is the thermal gluon mass. Note that here and in the following
we neglect masses of the particles at legs of the diagram. Since typical
scattering particles are thermal $p\sim k\sim T$, including them
would introduce mass-squared corrections to our collision integrals
which we can safely neglect. For the $SU(2)$ and $U(1)$
boson exchange diagrams for leptons
\eqn\mlepton{ |{\cal M}_W|^2=  A_W {s ^2\over (t - m_W^2 ) ^2 }
\qquad  |{\cal M}_B|^2=  A_B {s^2\over (t - m_B^2)^2}  }
where $A_W=36g_W ^4$, $A_B=78 g_w ^4 \tan^4 \theta_w$, $g_w$
the weak coupling constant, $\theta_w$ the Weinberg angle,
and $m_W$ and $m_B$ the thermal masses of the $SU(2)$ and $U(1)$ gauge
bosons respectively. These amplitudes include the counting over
all the fermions and anti-fermions which the given fermion can scatter off.
Since the $t$-channel processes are dominated by the infrared exchange momenta
we have used the same approximation to the boson propagator as in Appendix C
of {\JPTlonga}: we have assumed the exchange boson is Debye screened and hence
the propagator is corrected by $t\rightarrow t-m^2$,
where $m(T)$ is the corresponding Debye
mass. This approximation is reasonable for longitudinal bosons; but for
transverse bosons, since they are not Debye screened at the one-loop level, it
underestimates the scattering rates.

To see that the first and fourth integral are zero we simply
use the symmetry of the matrix element under exchange of
initial and final momenta i.e. $p \rightarrow p'$ and $k \rightarrow
k'$.

Using  parity invariance of the amplitudes
($\vec p_i\rightarrow -\vec p_i$)
%the rotational invariance
one can show that the
third and fifth integrals are also zero. We are left with
only two integrals which we denote as $I^0$ (the {\it second\/} integral)
and $I^3$ (the {\it sixth\/} integral)
which can be written in a Lorentz invariant form as follows:
\eqn\whatsleft{\eqalign{& \qquad I^0 = \int {{d\!\!\!^-}^3 p \over 2 p_o}
   \int   {{d\!\!\!^-}^3 k \over 2 k_o}
f_p f_k \iota ^0\,,\qquad\qquad\qquad
I^3 =  \int {{d\!\!\!^-}^3 p \over 2 p_o}
   \int   {{d\!\!\!^-}^3 k \over 2 k_o}
f_p f_k  \iota ^3 \cr
\iota ^0 & =
\int {{d\!\!\!^-}^3 p' \over 2 p'_o} (1-f_{p'})
       \int  {{d\!\!\!^-}^3 k' \over 2 k'_o} (1-f_{k'}) |{\cal M}|^2
    (2\pi)^4{\delta}^4 (p+k-p'-k')
{1\over 2}\big [ (p-p')\cdot u\big ]^2 \cr
\!\!\iota ^3 & \!\!= \!\!
\int {{d\!\!\!^-}^3 p' \over 2 p'_o} (1-f_{p'})
       \int  {{d\!\!\!^-}^3 k' \over 2 k'_o} (1-f_{k'}) |{\cal M}|^2
    (2\pi)^4{\delta}^4 (p+k-p'-k') {1\over 6}
\left [-t+ \{ (p-p')\cdot u\} ^2 \right ]
}}
where $u^\mu$ is the plasma vector which is in the plasma frame
$u^\mu=(1,\vec 0)$, and
$f_{p_i}=1/(1+\exp (p_i\cdot u/T))$ are  the population densities.
Dropping the Pauli blocking factors for the out-going states we can
evaluate the integrals over $p'$ and $k'$ in the center of mass frame.
In this frame $u^\mu=(\gamma,\gamma \vec v)$, which can be written in terms of
the plasma frame quantities as
$\gamma=(p^{pl}_0+k^{pl}_0)/\sqrt{s^{pl}}$,
$\gamma \vec v= (\vec p^{pl}+\vec k^{pl})/\sqrt{s^{pl}}$,
$s^{pl}=2p^{pl}\cdot k^{pl}$.
For convenience we choose  the axes such that $\vec p$ is along
the $z$-axis: $\vec p=p(0,0,1)$, $\vec v$ has {\it zero\/} azimuthal angle:
$\vec v = v (\sin\beta, 0 , \cos\beta)$ and
$\vec p'= p' (\sin\theta '\cos\phi ',\sin\theta '\sin\phi ',\cos\theta ')$.
After some algebra and keeping only the leading logarithm, which occurs
for small $\theta'$, we get
\eqn\gammaTv{\eqalign{\iota^0 & ={A\over 8\pi}(\gamma v p \sin\beta)^2
\ln{4p^2\over m_g^2} \cr
3\iota^3 & ={A\over 8\pi}\left [(\gamma v p \sin\beta)^2  +2p^2\right ]
\ln{4p^2\over m_g^2}
}}
where $A=2A_q$ for quarks and $A=A_W$, $A=A_B$ for the leptons.
In order to recast this in a Lorentz invariant form, note that the Lorentz
scalars we can construct from $u^\mu$, $p^\mu$ and $k^\mu$ are
$u\cdot p$, $u\cdot k$ and $p\cdot k$ (since $p\cdot p\simeq 0$,
$k\cdot k\simeq 0$, $u\cdot u=1$). The unique solution is
$(\gamma v p \sin\beta)^2=u\cdot p\; u\cdot k -{p\cdot k\over 2}$, so that we
have
\eqn\gammaTvi{\eqalign{\iota^0 & ={A\over 8\pi}
\left [ u\cdot p\; u\cdot k -{p\cdot k\over 2}\right ]
\ln{2p\cdot k\over m_g^2} \cr
3\iota^3 &
={A\over 8\pi}\left [u\cdot p\; u\cdot k +{p\cdot k\over 2}\right ]
\ln{2p\cdot k\over m^2}
}}
It is now possible to do the remaining integrations using
the approximation $\int f_x x^n\ln x dx \approx \ln (n+1) \int f_x x^n  dx$,
 $f_x=(1+exp\; x)^{-1}$;
we find
\eqn\resultq{\eqalign { I^0=I^3 &= { A \over 1024 \pi^5}
9\zeta_3^2 \ln{36 T^2\over m^2}  T^6
} }
Now going back to the definitions of $\Gamma_T$ and $\Gamma_v$ in
\collintexp\ and \collintb\ we have
\eqn\relations{ \Gamma_T= {1 \over 4 \rho_o T_o} I^o \qquad
                \Gamma_v ={ 3\over 4 \rho_o T_o} I^3   }
where $\rho_o ={ 21 \over 8 \pi^2}\zeta_4 T_o^4$.
For the quarks this gives, when we use $m_g^2=8\pi \alpha_s T_o^2
\approx 3.6 T_o^2$  as described in Appendix A of \JPTlonga\  for
$\alpha_s={1 \over 7}$ at $T_o\sim 100$GeV,
\eqn\relations{\Gamma_v = 3 \Gamma_T=
{18\zeta_3^2\over 7\zeta_4\pi} \alpha_s^2 \ln{9\over 2\pi\alpha_s} \; T_0
\approx {T_0\over 20}
}
Using the relation $D=b/3a\Gamma_v$ from
\diffusion, this gives us the quark diffusion constant
\eqn\diffconst{
 D_q^{-1} = {8\zeta_2\over\pi}\alpha_s^2\ln{9\over 2\pi\alpha_w}\; T_0
\approx {T_0 \over 5}
}
For the leptons we calculate
\eqn\diffconsti{\eqalign{
&  D_W^{-1}  = {9\zeta_2\over 2\pi}\alpha_w^2\ln{27\over 5\pi\alpha_w}\; T_0
\approx {T_0 \over 100} \cr
D_B^{-1} & = {39\zeta_2\over 4\pi}\alpha_w^2 \tan ^4\theta_W
\ln{27\over\pi\alpha_w\tan^2\theta_W}\; T_0 \approx {T_0 \over 290}
 }}
where we used
$m_B^2={4\pi \over 3}\alpha_w \tan^2 \theta_W T_0^2\simeq 0.04 T_0^2$
and $m_W^2={20\pi \over 3}\alpha_w \tan^2 T_0^2\simeq 0.7 T_0^2$, and taking
$\alpha_w={1 \over 30}$ and $\sin ^2 \theta_W=0.23$.

These values for the diffusion constants agree very well
with the values we obtained in Appendix A of \JPTlonga.
The methods employed differ in that the first allowed a more
general perturbation to the phase space density, but required
an assumption about the near equality
of the energy of ingoing and outgoing scattered particles.
The method presented here is based on a more restrictive
form for the distribution function, but involves no
additional approximations.

\medskip

\centerline {\bf Appendix B. Collision Integrals for Decay Processes }

In this appendix we evaluate
the gluon exchange helicity-flip rate which
dominates the hypercharge violating interactions.
The relevant Feynman diagram is shown in Figure 2.
The top/Higgs helicity-flip processes (hypercharge
violating Higgs exhange and Higgs absorption or emission)
are  slower because
there are fewer particles to scatter off.

The rate for a top quark of helicity
$\lambda$ to scatter into one of helicity $\lambda'$ is
\eqn\appxCfliprate{\eqalign{
\Gamma_{\lambda \rightarrow \lambda'}= &
{12\over T^3}\int {{d\!\!\!^-}^3 p \over 2 p_o}f_o(p_o/T)
       {{d\!\!\!^-}^3 k \over 2 k_o}f_o(k_o/T) {\cal I}_f\cr
{\cal I}_f= &
\int{{d\!\!\!^-}^3 p' \over 2 p'_o}{{d\!\!\!^-}^3 k' \over 2 k'_o}
{\delta\!\!\!^-}^4 (p+k-p'-k') |{\cal
M}_{\lambda\rightarrow\lambda'}|^2
}}
on the wall where particles acquire a mass. The
factor $12/T^3=1/3n_0a$ is
in accord with the rate definition $\Gamma_f$ in the  diffusion
equations
\coreeqns.
Even though the masses are spatially dependent, for our purposes
it is sufficient to evaluate the rate assuming a constant mass.
The corrections to this approximation are of order $l\nabla m$ where
 $l\sim 1/M_g$ is the Debye screening length of the gluon so that
$l|\nabla m|/m\sim 1/g_s TL<<1$.
In \appxCfliprate\ we consider an in-going fermion with
mass $m_1$ momentum and helicity
$\{ p\, ,\lambda\}$
which scatters off a fermion $m_2$, $k$ (with arbitrary helicity) into
$\{ p'\, ,\lambda'\}$, $k'$
with the appropriate spin dependent scattering amplitude
${\cal M}_{\lambda\rightarrow\lambda'}$. The notation we use in
\appxCfliprate\
is familiar to the reader: $n_0=3\zeta_3T^3/4\pi^2$, $\zeta_3=1.202$,
$f_o(x)=1/(\exp x+1)$, ${d\!\!\!^-}^3 k\equiv d^3k/(2\pi)^3$.
We ignore Pauli blocking factors.

In order to evaluate $|{\cal M}_{\lambda \rightarrow \lambda'}|^2$
we define,
 following e.g.
\ref\Haber{H. E. Haber, preprint SCIPP 93/49,
NSF-ITP-94-30 (April 1994),  to appear in the proceedings of the
21st SLAC Summer Institute on Particle Physics: {\it Spin Structure in
High Energy Processes}, SLAC, Stanford, 26 July - 6 August, 1993.},
the spin-four vector as
\eqn\appxCspinvector{s^\mu ={2 \lambda\over m} \left ( |\vec p\,|,\,
p_0
{\vec p\over |\vec p\,|}
\right )\qquad \qquad \lambda=\pm {1\over 2}
}
where $\lambda$  is the  helicity, so that $s\cdot p=0$ and
$s\cdot s=-1$. The
helicity projection operators for a massive spin-$1/2$ particle
eigenspinor $u(p,\lambda)$  are
\eqn\appxCspinorprojection{u(p,\lambda)\bar u(p,\lambda)=
{1\over 2}(1+\gamma_5 s\!\!\!/\,) (p\!\!\!/ \, +m)
}
and similarly for anti-particles.

The scattering amplitude squared reads
\eqn\appxCamplitudes{\eqalign{
|{\cal M}_{\lambda \rightarrow \lambda'}|^2= &
{2A_s\over \bigl [ (p-p')^2-M_g^2  \bigr ]^2}
{1\over 16} T_1\cdot T_2\cr
T_1^{\nu\mu}= & {\rm Tr\,}\bigl [(k\!\!\!/ +m)\gamma^\nu
(k'\!\!\!\!/ +m)\gamma^\mu   \bigr ]\cr
{T_2}_{\,\nu\mu}= & {\rm Tr\,}\bigl [(p\!\!\!/ +m)\gamma_\nu
{1\over 2}\bigl (1+\gamma_5 s\!\!\!/\,(p',\lambda')\bigr )
(p'\!\!\!\!/ +m)\gamma_\mu
{1\over 2}\bigl (1+\gamma_5 s\!\!\!/\,(p,\lambda)\bigr ) \bigr ]\cr
}}
where $A_s=32 g_3^4= 512\pi^2\alpha_s^2$ ($\alpha_s=g_3^2/4\pi=1/7$)
includes the 12 quarks  and anti-quarks that a quark can
scatter off via the gluon exchange.
Rather than evaluating
$|{\cal M}_{\lambda \rightarrow \lambda'}|^2$ for each pair
$\{\lambda\, ,\lambda'\}$ we can
define the helicity-flip amplitude
\eqn\appxCamplitudesi{\eqalign{|{\cal M}_{-}|^2={1\over 2}
\sum_{\lambda'=-\lambda}
|{\cal M}_{\lambda \rightarrow \lambda'}|^2
}}
and the no-flip amplitude
\eqn\appxCamplitudesii{\eqalign{|{\cal M}_{+}|^2={1\over 2}
\sum_{\lambda'=\lambda}
|{\cal M}_{\lambda \rightarrow \lambda'}|^2
}}
which simplifies the problem considerably.
We still have to plough through  rather lengthy algebra to arrive at an
expression for $T_1\cdot T_{2\,\pm}$
\eqn\appxCToneTtwo{\eqalign{
{1\over 16}T_1\cdot T_{2\;\pm}= &
{1\over 2}\sum_{\lambda'=\pm\lambda}{1\over 16}T_1\cdot T_{2}=\cr
& (p\cdot k \, p'\cdot k' +
p\cdot k' \, p'\cdot k -
m_1^2\, k\cdot k'- m_2^2 \, p\cdot p' +2 m_1^2 m_2^2)\cr
\pm {1\over 2}\sum_{\lambda'=\lambda} & \bigl [
-s\cdot s'\, (p\cdot k \, p'\cdot k' +   p\cdot k' \, p'\cdot k
-p\cdot p' \, k\cdot k')\cr
& -p\cdot p'\, (k\cdot s \, k'\cdot s' +   k\cdot s' \, k'\cdot s )
+p\cdot k \, p'\cdot s \, k'\cdot s' \cr
& +p\cdot k'\, p'\cdot s \, k\cdot s' +
p'\cdot k\, p\cdot s' \, k'\cdot s \cr
& +p'\cdot k'\,  k\cdot s \, p\cdot s'
-k\cdot k' \, p\cdot s' \, p'\cdot s \cr
& +m_1^2 \, ( k\cdot s \, k'\cdot s' +   k\cdot s' \, k'\cdot s -
m_2^2\, s\cdot s' )
\bigr ]\cr
}}
where  $s_\mu=s_\mu (p,\lambda)$
and $s'_\mu=s_\mu (p',\lambda')$.

Since this form is Lorentz invariant,  the integration over
the out-going momenta  $p'$ and $k'$ is rather straightforward
in the center of mass frame: $\vec p+\vec k=0={\vec p}'+{\vec k}'$.
To leading order in  $m_1^2$  we obtain
\eqn\appxCIpm{\eqalign{
{\cal I}_{-}= & {A_s\over 8\pi}{m_1^2\over p\cdot k}
\ln\biggl [
1+{2 p\cdot k\over M_g^2}\biggr ]\cr
{\cal I}_{+}= & {A_s\over 2\pi}{p\cdot k\over M_g^2}
}}
with  the  gluon thermal mass squared $M_g^2=8\pi\alpha_s T^2$.

We can now integrate both ${\cal I}_{-}$ (in the leading
logarithm approximation as explained in appendix B of \JPTlonga)
and  ${\cal I}_{+}$ according to \appxCfliprate\
to obtain the helicity-flip rate $\Gamma_{-}$ and the no-flip
rate $\Gamma_+$
\eqn\appxCrates{\eqalign{
\Gamma_f= &\Gamma_-=
{24\ln^2 2\over \pi^3}{m_1^2\over T^2}\alpha_s^2\iota T \cr
\Gamma_+= & {54\zeta_3^2\over \pi^4}\alpha_s T\cr
}}
where  $\iota=\int_0^2{dz\over z}\ln(1+x_1^2 z/2\pi\alpha_s)\simeq
\zeta_2+(1/2)\ln^2 (x_1^2/\pi\alpha_s)-\pi\alpha_s/x_1^2\simeq 2.5$
is the angular integral ($z=p\cdot k/pk$) which is approximately
unity;
$x_1\sim 1.3$ is the value of the momentum $p/T$ at which
the momentum integrals in \appxCfliprate\ peak.

We use this estimate of $\Gamma_f$ in
Section 6.
Using the same method one could evaluate the
helicity-flip $W$- and $B$-exchange processes relevant for the leptons.
However since our
main focus in this work is
on top quark mediated  baryogenesis, we shall not do so here.

\medskip

\centerline {\bf Appendix C. Finite Temperature Dispersion Relation}

In this appendix we follow Appendix B of \JPTlonga\
to arrive at a  finite temperature Dirac equation in momentum space.
In the presence of a $Z$ field condensate it is convenient to write
the equation in terms of $2\times 2$ chiral spinors in the plasma frame
\eqn\appxBplasmaframeequations{\eqalign{\left [ (E- g_A Z_0+c_R)-
\vec\sigma\cdot (\vec P- g_A \vec Z )\right ]\Psi_R +m_R\Psi_L & =0 \cr
m_L\Psi_R+ \left [ (E+ g_A Z_0+c_L)+
\vec\sigma\cdot (\vec P+ g_A \vec Z )\right ]\Psi_L & =0\cr
}}
where the notation is that of \JPTlonga.
We will consider a planar wall moving in the positive $z$-direction with
velocity $v_w$
with a pure gauge condensate $Z^\mu=(Z_0,0,0,Z_z)(z)$ so that $Z_0=-v_w Z_z$.
The dispersion relation we require is obtained by setting the determinant of
\appxBplasmaframeequations\ to {\it zero}:
\eqn\appxBdispersionrelation{\eqalign{& \left [
 (E+c)^2-{\vec P}^2-m_T^2 -\left (g_A Z_0+{1\over 2}\Delta c \right )^2
+g_A^2 Z_z^2
\right ]^2= \cr
& 4 {\vec P}^2 \left (g_A Z_0+{1\over 2}\Delta c \right )^2
+4 g_A^2 Z_z^2 \left [ (E+c)^2 -{\vec P}_\perp^2
\right ] -8 P_z g_A Z_z(E+c)\left (g_A Z_0+{1\over 2}\Delta c\right ) \cr
}}
where we have defined
\eqn\appxBdefinitionCdelC{c={c_R+c_L\over 2}\,,\qquad
\Delta c=c_L-c_R\,,\qquad
m_T^2= {m^2\over (1+a_L)(1+a_R)}
}
The dispersion relation for anti-particles is obtained by
the replacement $P_z\rightarrow -P_z$ and $Z_0\rightarrow -Z_0$
(recall that $Z_z$ is even under $CP$).

In the limit when both $Z_0, Z_z\rightarrow 0$ we find for particles
\eqn\appxBdispersionnofield{(E+c)^2=
\left (|\vec P|\pm {1\over 2} \Delta c\right )^2 +m_T^2
}
which reduces in the high momentum limit to
$E^2 = P^2 + m_T^2 +2M_{L,R}^2$.

A second case which is easy  to solve is when $Z_z=0$
 and $Z_0\neq 0$. This is  relevant to the case of
spinodal decomposition (second order phase transition):
\eqn\appxBdispersiononlyZo{(E+c)^2= \left [
|\vec P|\mp \left (g_A Z_0+{1\over 2}\Delta c\right )
\right ]^2 +m_T^2
}
which in the high momentum limit simplifies to
\eqn\appxBdispersiononlyZohiK{E^2= \left [
|\vec P|\mp g_A Z_0 \right ]^2 +  2M_{L,\, R}^2 +m^2
}
Therefore, as long as $P>>M_{L,\, R}$, the motion of a thermal excitation
is not affected (to leading order) by the dispersive plasma effects.
In a first order phase transition this is not the relevant case since
$Z^\mu$ is space like;
in the plasma frame for example $ |Z_0/Z_z|=v_w$ so for slow walls it
makes sense to neglect $Z_0$.

To make a detailed comparison with the free particle case
we now proceed to a more systematic study of   {\appxBdispersionrelation}
and rewrite it as follows
\eqn\appxBdispersionrelationi{\eqalign{& \lambda=E+c\,,\cr
& 0= \lambda^4-2 \left [ \vec P^2+m_T^2+g_A^2 Z_z^2
 +\left (g_A Z_0+{1\over 2}\Delta c \right )^2
\right ] \lambda ^2 +
 8 P_z g_A Z_z\left (g_A Z_0+{1\over 2}\Delta c\right ) \lambda +\cr
&\left [
\vec P^2+m_T^2-g_A^2 Z_z^2 + \left (g_A Z_0+{1\over 2}\Delta c\right )^2
\right ]^2
\!\!\! + 4m_T^2\left [
 \left (g_A Z_0+{1\over 2}\Delta c\right )^2-g_A^2 Z_z^2
\right ]
- 4 P_z^2 g_A^2 Z_z^2
}}
We now consider a weak $Z$ field expansion of this equation
and write the solution in the form:
\eqn\appxBdispersionrelationsolution{\eqalign{ \lambda=\lambda_0 (1+\epsilon)
}}
where $\lambda_0$ is the solution of \appxBdispersionrelationi\
with linear term in $\lambda$ neglected:
\eqn\appxBdispersionrelationlambdao{\eqalign{ \lambda_0^2 & =
e_0^2\pm \sqrt\Delta_0 \cr
e_0^2 & = \vec P^2 +m_T^2 +g_A^2 Z_z^2 +
          \left (g_A Z_0+{1\over 2}\Delta c\right )^2 \cr
\Delta_0 & = 4 \left ( \vec P^2+g_A ^2 Z_z ^2 \right )
\left (g_A Z_0+{1\over 2}\Delta c\right )^2 +
4 g_A^2 Z_z^2 ( m_T^2+P_z^2 )
}}
The leading order correction $\epsilon$  reads
\eqn\appxBdispersionrelationepsilon{ \epsilon\simeq
\mp{2 g_A Z_z P_z \left (g_A Z_0+{1\over 2}\Delta c\right ) \over
\lambda_0 \sqrt \Delta_0
}}
where the  signs coincide with {\appxBdispersionrelationlambdao}.
In order to make more transparent what the dispersion relation
\appxBdispersionrelationsolution\ -- \appxBdispersionrelationepsilon\
mean we look at its high momentum limit.

We now restrict ourselves to the case $Z_0=0, Z_z \neq 0$, which
describes the case of a wall at rest in the plasma. Since
for a stationary wall profile $Z_0=-v_w Z_z$, we anticipate
that this will be a good approximation for a sufficiently slow wall.
There are two cases to consider depending on
which term in the determinant $\Delta_0$  in
{\appxBdispersionrelationlambdao\/} dominates
\eqn\appxBdeltao{\eqalign{\sqrt \Delta_0 \simeq
   2g_A Z_z\sqrt {P_z^2+m_T^2}\,,\qquad & {\rm for}  \quad
   (\Delta c)^2 P^2 << 4 g_A^2 Z_z^2 (P_z^2+m^2)\cr
 \sqrt \Delta_0 \simeq |\Delta c| P\simeq (M_L^2-M_R^2){E\over P}
\,,\qquad & {\rm for}  \quad (\Delta c)^2 P^2 >> 4 g_A^2 Z_z^2(P_z^2+m^2)\cr
}}
In the high momentum limit these two cases reduce to a comparison
between $M_L^2-M_R^2$ and $2 g_A Z P_z$. Since $M_L^2-M_R^2 \approx
\alpha_w T^2$ (see \JPTlonga\ ), and we can write $2g_A Z\equiv
\Theta_{CP}/L$,
the two cases become approximately $\Theta_{CP} >  \alpha_w LT$ and
$\Theta_{CP} <  \alpha_w LT$, which we therefore refer to as
cases of  `strong'  and `weak' condensates respectively.
(Note that for a realistic wall thickness in order to be in the `strong'
regime $\Theta_{CP}$ must be of order {\it unity\/}.)

In the first case the dispersion relations are
\eqn\appxBdispersionrelationparticles{\eqalign{ (E^{L,\,R}+c)^2 & =
P_\perp^2+\left ({\rm sign}\, P_z\, \sqrt{P_z^2+m_T^2}\pm g_AZ_z\right )^2
+\left ({\Delta c/2}\right )^2 \cr
\mp & ({\Delta c/ 2}){|P_z|\over\sqrt{P_z^2+m_T^2}}
2 \left (E^{L,\, R}+c\right )
+ \left ({\Delta c/2}\right )^2{P_z^2\over P_z^2+m_T^2}
 \cr
\left ( E^{\bar L,\, \bar R}+c\right )^2 & =
P_\perp^2+\left ({\rm sign}\, P_z\, \sqrt{P_z^2+m_T^2}\mp g_AZ_z\right )^2
+\left ({\Delta c/2}\right )^2  \cr
\mp & ({\Delta c/2}){|P_z|\over\sqrt{P_z^2+m_T^2}}
2 \left (E^{\bar L,\, \bar R}+c\right )
+ \left ({\Delta c/2}\right )^2{P_z^2\over P_z^2+m_T^2}.
 \cr
}}
We have adopted here the notation used in the main text - labelling
the WKB states by the chiral states they deform into in the
symmetric phase. A careful inspection of these relationships in the high
momentum limit $P>>M_{L,\, R}$, reveals that they reduce, to leading order in
coupling constants, to the zero temperature dispersion relations {\it plus}
the finite temperature mass $\sqrt{2}M_L$ for the left handed particles and
their anti-particles  and $\sqrt{2}M_R$ for the right handed particles and
their anti-particles.

Firstly we note that in the free case
(when we set $T=0$) \appxBdispersionrelationparticles\ reproduces
precisely the dispersion
relations we had in \dispersion\ and \dispersionap.
Secondly, when $Z_z=0$ we
see that the effect of the thermal corrections
is to split the left-handed and right-handed states (by an amount
$\propto \Delta c$),
but to leave the degeneracy of particles and anti-particles
intact. Finally, and most importantly,
 we see that the corrections to the free dispersion relation
are small in this case $\Theta_{CP} > \alpha_w LT$, so that the analysis in the
main text applies.

The second case of a `weak' condensate in \appxBdeltao\
gives dispersion relations
\eqn\appxBdispersionrelationparticlesi{\eqalign{
\left (E^{L,\,R}+c\right )^2 & =
\left (P\mp\Delta c/2\right )^2+m_T^2
\pm 2 g_A Z_z {P_z\over P}\sqrt{(P\mp \Delta c/2)^2+m_T^2}\cr
\left (E^{\bar L,\,\bar R}+c\right )^2 & =
\left (P\pm \Delta c/2 \right )^2+m_T^2
\mp 2g_A Z_z {P_z\over P} \sqrt{(P\pm \Delta c/2)^2+m_T^2}\cr
}}
It is instructive to re-write the first relation in the limit when
$P>>M_{L,\, R}$
\eqn\appxBdispersionrelationparticlesii{
\left (E^{L,\,R}\right )^2  \simeq P^2  +m^2 + 2 M_{L,\, R}^2
\pm  2 g_A Z_z {P_z\over P} \sqrt{P^2+m^2}
}
When  $m=0$ there should be  no physical effect
due to the $Z$ field since it is then just pure
gauge. We see that this is the case  since
the dispersion relation can then be
written  $\left (E^{L,\,R}+c\right )^2
=\left (P'\mp\Delta c/2\right )^2$ where $P'
=|\vec {P'}|$ and $\vec{P'}=(P_\perp, P_z \pm g_A Z_z)$
(compare this form with {\appxBdispersiononlyZo}).
When $m \neq 0$ however, just as in the zero temperature
case, the dispersion relation \appxBdispersionrelationparticlesi\
leads to  a non-zero acceleration:
$\dot v_z=\mp\partial_z (g_AZ_z m^2)(P_z^3/P^3 E^3)
-\partial_zm^2/2E^2$,
where we have assumed that $E$ and $P_\perp$ are
conserved in the plasma frame, an approximation correct to leading
order in $v_w$.
Comparing this result with \acc\ we
see that, in this second case of a `weak' condensate,
the force term has the same form ($\propto \partial_z (g_AZ_zm^2)$)
but a somewhat different momentum dependence.
Corresponding corrections to the analysis presented in the text
would be required to describe this case, which we
anticipate would lead to minor numerical changes in
the coefficients of the terms in the fluid equations
if the analysis is carried through in the same way.

\listrefs

\figures
\fig{1}{Dispersion relation for particles in a background
axially coupled pure  gauge field. The plots show the energy $E$
as a function of canonical momentum $p_z$ for $p_\perp=0$,
for (i) $m > g_A Z$, (ii) $m <  g_A Z$ and (iii) $m=0$.
 }
\fig{2}{ Vector boson exchange diagrams which dominate
in damping temperature and velocity perturbations in the
fluid, and play a leading role
in determining the diffusion properties of different
particle species. The gluon exchange diagram also,
in the presence of a quark mass term,
describes the key hypercharge-violating process, namely
the helicity flip process computed in Appendix B.
}
\fig{3}{ Chirality flipping Higgs process which contributes to
the damping of a chiral chemical potential in front of the
bubble wall.
}
\fig{4}{ Solutions for the chiral chemical potential $\mu$ in the
background of a bubble wall, with a `ramp' ansatz for the
$CP$ violating condensate field $m^2 Z$.
The solutions for ${\bar \mu \over T_o}$ are sketched
for the two cases (i) ${v_w L \over D}<< 1$ and (ii) ${ v_w L \over
D}>>1$.
}
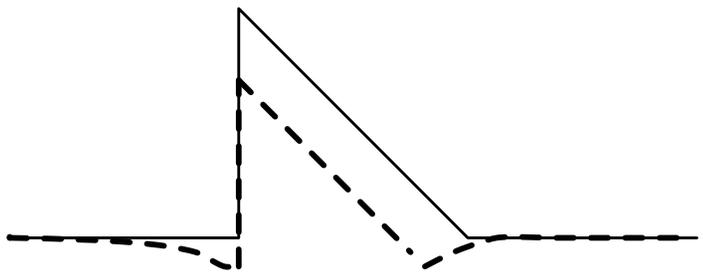
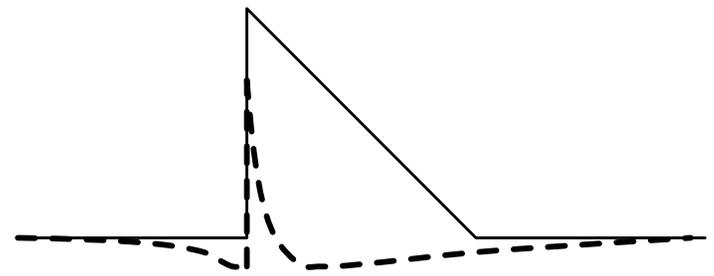
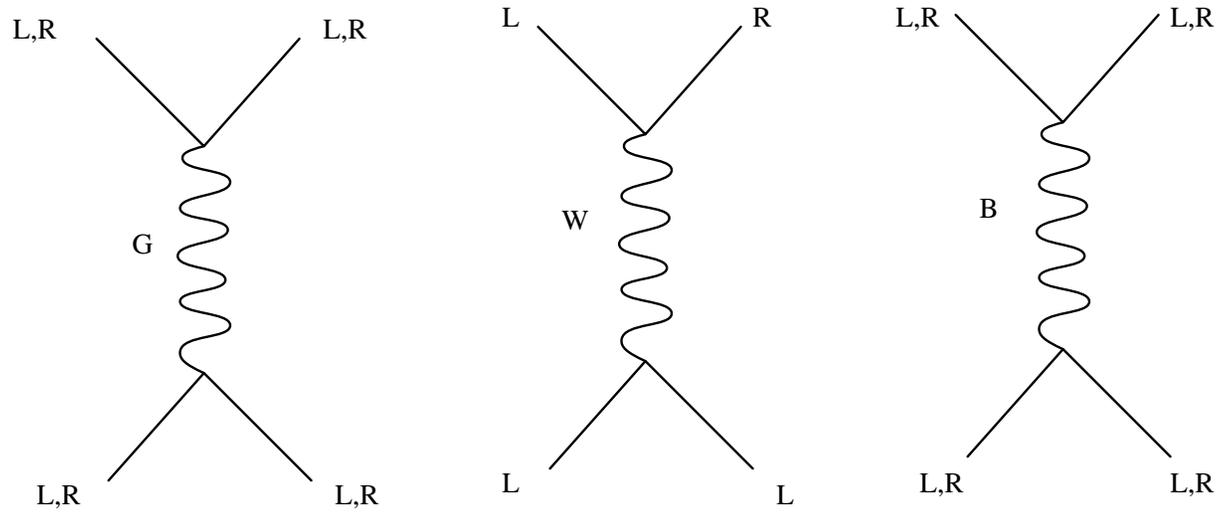
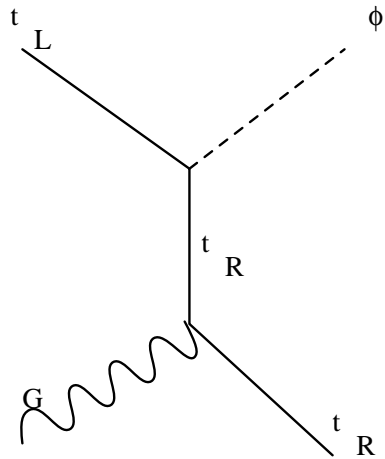
\bye